\documentclass[%
12pt
]
{article}

\usepackage{ifthen}
\usepackage{afterpage}
\usepackage{cite}

\title{One-loop RGEs for two left-right SUSY models}	
\author{N. Setzer \& S. Spinner}	
\newcommand{\up}[1]{\textsuperscript{#1}}
\newcommand{\deriv}[2]{   { d {#1} \over d {#2} }  }

\newcommand{\conj}[1]{{#1}^*}
\newcommand{\half}{\frac{1}{2}}
\newcommand{\third}{\frac{1}{3}}

\newcommand{\sixth}{\frac{1}{6}}

\newcommand{\eighth}{\frac{1}{8}}
\newcommand{\twoByTwo}[4]%
		{%
		\left(%
		\begin{array}{cc}%
		#1	& #2	\\
		#3	& #4
		\end{array}%
		\right)%
		}
		
\newcommand{\threeByThree}[9]%
		{%
		\left(%
		\begin{array}{ccc}%
		#1	& #2	& #3	\\
		#4	& #5	& #6	\\
		#7	& #8	& #9
		\end{array}%
		\right)%
		}

\newcommand{\colTwoVect}[2]{  \inp{ \begin{array}{c} #1 \\ #2 \end{array} }  }

\newcommand{\rowTwoVect}[2]{  \inp{ \begin{array}{cc} #1 & #2 \end{array} }  }

\newcommand{\SUTwoGen}[2][2]%
{%
\ifthenelse{\equal{#1}{3}}%
	{%
	\ifthenelse{\equal{#2}{0}}%
		{%
		\left(%
		\begin{array}{ccc}%
		1	& 0	& 0	\\
		0	& 1	& 0	\\
		0	& 0	& 1
		\end{array}%
		\right)%
		}%
		{%
		\ifthenelse{\equal{#2}{1}}%
			{%
			\left(%
			\begin{array}{ccc}%
			0			& \frac{1}{\sqrt 2}	& 0			\\
			\frac{1}{\sqrt 2}	& 0			& \frac{1}{\sqrt 2}	\\
			0			& \frac{1}{\sqrt 2}	& 0
			\end{array}%
			\right)%
			}%
			{%
			\ifthenelse{\equal{#2}{2}}%
				{%
				\left(%
				\begin{array}{ccc}%
				0			& - \frac{i}{\sqrt 2}	& 0			\\
				\frac{i}{\sqrt 2}	& 0			& - \frac{i}{\sqrt 2}	\\
				0			& \frac{i}{\sqrt 2}	& 0
				\end{array}%
				\right)%
				}%
				{%
				\left(%
				\begin{array}{ccc}%
				1	& 0	& 0	\\
				0	& 0	& 0	\\
				0	& 0	& -1
				\end{array}%
				\right)%
				}%
			}%
		}%
	}%
{
\ifthenelse{\equal{#1}{4}}%
	{%
	\ifthenelse{\equal{#2}{0}}%
		{%
		\left(%
		\begin{array}{cccc}%
		1	& 0	& 0	& 0	\\
		0	& 1	& 0	& 0	\\
		0	& 0	& 1	& 0	\\
		0	& 0	& 0	& 1
		\end{array}%
		\right)%
		}%
		{%
		\ifthenelse{\equal{#2}{1}}%
			{%
			\left(%
			\begin{array}{cccc}
		0			& \sqrt{\frac{3}{10}}	& 0			& 0			\\
		\sqrt{\frac{3}{10}}	& 0			& \sqrt{\frac{2}{5}}	& 0			\\
		0			& \sqrt{\frac{2}{5}}	& 0			& \sqrt{\frac{3}{10}}	\\
		0			& 0			& \sqrt{\frac{3}{10}}	& 0
			\end{array}%
			\right)%
			}%
			{%
			\ifthenelse{\equal{#2}{2}}%
				{%
				\left(%
				\begin{array}{cccc}%
		0			& -i\sqrt{\frac{3}{10}}	& 0			& 0				\\
		i \sqrt{\frac{3}{10}}	& 0			& -i \sqrt{\frac{2}{5}}	& 0				\\
		0			& i \sqrt{\frac{2}{5}}	& 0			& -i \sqrt{\frac{3}{10}}	\\
		0			& 0			& i \sqrt{\frac{3}{10}}	& 0
				\end{array}%
				\right)%
				}%
				{%
			\left(%
			\begin{array}{cccc}%
			\frac{3}{\sqrt 10}	& 0			& 0			& 0	\\
			0			& \frac{1}{\sqrt 10}	& 0			& 0	\\
			0			& 0			& -\frac{1}{\sqrt 10}	& 0	\\
			0			& 0			& 0			& - \frac{3}{\sqrt 10}
				\end{array}%
				\right)%
				}%
			}%
		}%
	}%
	{%
	\ifthenelse{\equal{#2}{0}}%
		{%
		\left(%
		\begin{array}{cc}%
		1	& 0	\\
		0	& 1
		\end{array}%
		\right)%
		}%
		{%
		\ifthenelse{\equal{#2}{1}}%
			{%
			\left(%
			\begin{array}{cc}%
			0	& 1	\\
			1	& 0
			\end{array}%
			\right)%
			}%
			{%
			\ifthenelse{\equal{#2}{2}}%
				{%
				\left(%
				\begin{array}{cc}%
				0	& - i	\\
				i	& 0
				\end{array}%
				\right)%
				}%
				{%
				\left(%
				\begin{array}{cc}%
				1	& 0	\\
				0	& -1	
				\end{array}%
				\right)%
				}%
			}%
		}%
	}%
}
}
	
\newcommand{\inp}[2][0cm]{ \left( #2 \parbox[h][#1]{0cm}{} \right) }
\newcommand{\inb}[2][0cm]{ \left[ #2 \parbox[h][#1]{0cm}{} \right] }

\newcommand{\nop}[1]{\left.{#1}\right.}
\newcommand{\real}[1]{\, \mbox{Re} \ifthenelse{\equal{#1}{^}}{^}{\, #1} }
\newcommand{\imag}[1]{\, \mbox{Im} \ifthenelse{\equal{#1}{^}}{^}{\, #1} }

\newcommand{\field}[1]{#1}
\newcommand{\superp}[1]{\tilde{#1}}

\newcommand{\Tr}{\,\mbox{Tr}\,}

\newcommand{\hide}[1]{}
		
\newcommand{\YPhimnYPhimn}[3][\beta]{%
		  \Tr\inp{3 h_{#2}^\dagger h_{#3} + h_{#2}^{\prime\dagger} h_{#3}^\prime}
		+ 4 \inp{ \mu_\Phi^{#1 \, \dagger} \mu_\Phi^{#1} }_{{#2}{#3}}%
		  	}
\newcommand{\YPhimnYPhimnTranspose}[3][\beta]{%
		  \Tr\inp{3 h_{#3} h_{#2}^\dagger + h_{#3}^\prime h_{#2}^{\prime\dagger} }
		+ 4 \inp{ \mu_\Phi^{#1} \mu_\Phi^{#1 \, \dagger}  }_{{#3}{#2}}%
		  	}
\newcommand{\YSmnYSmn}[2]{%
		    3 \mu_\Delta^{#1 \, *} \mu_\Delta^{#2}
		  + 3 \mu_{\Delta^c}^{#1 \, *} \mu_{\Delta^c}^{#2}
		  + 8 \Tr \inp{ \mu_\Phi^{#1 \, \dagger} \mu_\Phi^{#2} }
		  + \half \conj{\inp{Y^{#1 \mu \nu}}} Y^{#2 \mu \nu}
		  	}

\newcommand{\YPhimnhPhimn}[3][\beta]%
{%
		    \Tr\inp{6  h_{#2}^\dagger A_{Q {#3}} + 2 h_{#2}^{\prime\dagger} A_{L {#3}} }
		+ 8 \inp{ \mu_\Phi^{#1 \, \dagger} A_\Phi^{#1}  }_{{#2} {#3}}%
}
\newcommand{\YPhimnhPhimnTranspose}[3][\beta]%
{%
		    \Tr\inp{6 A_{Q {#3}} h_{#2}^\dagger  + 2 A_{L {#3}} h_{#2}^{\prime\dagger} }
		+ 8 \inp{ A_\Phi^{#1}  \mu_\Phi^{#1 \, \dagger} }_{{#3} {#2}}%
}

\newcommand{\YSmnhSmn}[2]{%
		    6 \mu_{\Delta}^{#1 \, *} A_{\Delta}^{#2}
		+ 6 \mu_{\Delta^c}^{#1 \, *} A_{\Delta^c}^{#2}
		+ 16 \Tr \inp{  \mu_\Phi^{#1 \, \dagger} A_\Phi^{#2}  }
		+ \conj{\inp{Y^{#1 \mu \nu}}} A_S^{#2 \mu \nu}%
		  	}
\newcommand{\YSmnbmn}[1]{%
		    6 \mu_\Delta^{#1 \, *} B_{\Delta}
		  + 6 \mu_{\Delta^c}^{#1 \, *} B_{\Delta^c}
		  + 16 \Tr \inp{ \mu_\Phi^{#1 \, \dagger} B_\Phi }
		  + \conj{\inp{Y^{#1 \mu \nu}}} B_S^{\mu \nu}%
		  	}
\newcommand{\massSum}[1][m]
{%
\ifthenelse{\equal{#1}{S}}%
	{%
	{\cal S}_3
	}%
	{%
	  4 \Tr\inp{	  m_Q^2
			- m_{Q^c}^2
			- m_L^2
			+ m_{L^c}^2
		 }
	+ 12 \inp{	  m_\Delta^2
			- m_{\bar{\Delta}}^2
			- m_{\Delta^c}^2
			+ m_{\bar{\Delta}^c}^2
		  }
	}%
}

\begin{document}             


\maketitle

\begin{abstract}
In this paper we present the renormalization group equations to one-loop order for all the parameters of two supersymmetric left-right theories that are softly broken.  Both models are based upon the gauge group $SU(3)^c \times SU(2)_L \times SU(2)_R \times U(1)_{B - L}$ and both contain an arbitrary number of bidoublets as well as singlets; however, one model uses doublets to break $SU(2)_R$ and the other uses triplets. 
\end{abstract}

\section{Introduction}
The recent discovery of oscillating neutrinos (implying that neutrinos are massive) has created definitive experimental evidence of a flaw in the standard model.  
This flaw can be rectified by adding an $SU(2)_R$ group to the standard model group structure.  This will allow for a Dirac mass term for the neutrinos and will also provide a Majorona mass term for the right handed neutrino via the seesaw mechanism 
\cite{Topic:Seesaw} 
when $SU(2)_R$ is broken.  These extensions are called left-right 
\cite{Journal:Phys.Rev.D10.275.1974, Journal:Phys.Rev.D11.2558.1975, Journal:Phys.Rev.D12.1502.1975, Journal:Nucl.Phys.B153.334.1979, Journal:Phys.Rev.D23.165.1981, Journal:Prog.Theor.Phys.67.1569.1982, ArXiv:hep-ph/0503115, Journal:Phys.Rev.D59.116004.1999}
 and they can also be imbedded in supersymmetry.  With SUSY, the models are dubbed SUSYLR
\cite{Journal:Phys.Rev.D58:115007.1998, ArXiv:hep-ph/9708492, Journal:Phys.Atom.Nucl.61.963-974.1998, ArXiv:hep-ph/0409216, Journal:Eur.Phys.J.C14.547-552.2000, Journal:Pramana.55.137-149.2000, ArXiv:hep-ph/9809370} 
and contain the attractive features of the supersymmetric standard model (e.g. providing a solution for the hierarchy problem and allowing for gauge coupling unification \cite{ArXiv:Martin}).  SUSYLR models have the additional appealing characteristics of solving the strong CP problem 
\cite{Journal:Phys.Rev.Lett.76:3490.1996, Journal:Phys.Rev.Lett.76:3486.1996, Journal:Phys.Rev.Lett.79:4744.1997, Journal:Phys.Rev.D54.3377-3381.1996, Journal:Phys.Rev.D65.056002.2002, Journal:Phys.Rev.D61.091701.2000}
, asymptotic parity invariance and automatic R-Parity Conservation 
\cite{Journal:Phys.Rev.D34.3457.1986, Journal:Phys.Lett.B228.79.1989, Journal:Phys.Rev.D46.2769.1992, Journal:Phys.Rev.D62.055010.2000, Journal:Phys.Rev.D48.4352-4360.1993, ArXiv:hep-ph/9803461}.

The parameters in these models are written down at a high scale of new physics such as the GUT or Planck scale.  In order to make predictions at lower energy levels, renormalization group equations (RGEs) must be calculated for these parameters, and their values extrapolated to the energy realm of current experiments.  In this paper the RGEs for two instances of left-right models are presented (non-SUSY Triplet left-right model equations can be found in \cite{Journal:Nucl.Phys.B358:181.1991}).  These equations were calculated to one loop order using the general N=1 supersymmetry RGEs given in \cite{Journal:Phys.Rev.D:SPMMTV:2LRGEs, Journal:Z.Phys.C30:247.1986} 
and agree (after accounting for the absence of $SU(2)_R$) with the subset previously published in \cite{Journal:Phys.Rev.D67.076006.2003}.  The following equations represent a completion and extension of those RGEs, and provide
a valuable tool for extrapolating down from higher scale physics to the scale of $SU(2)_R$ breaking
--
 at which point the model contains the MSSM and all couplings of interest can be extrapolated using the RGEs of the MSSM found in \cite{Journal:Phys.Rev.D:SPMMTV:2LRGEs}.

The two models used in this paper differ in their $SU(2)_R$ breaking fields: one uses $SU(2)_R$ doublets and the other $SU(2)_R$ triplets.  Theoretical consequences of these models can be found in various papers including 
\cite{Journal:Phys.Rev.D58:115007.1998, ArXiv:hep-ph/9708492, Journal:Phys.Atom.Nucl.61.963-974.1998, ArXiv:hep-ph/9809370, ArXiv:KSBabu.BDutta.RNMohapatra:SolveStrongCP}
 and 
\cite{ArXiv:hep-ph/0409216, ArXiv:KSBabu.BDutta.RNMohapatra:PartialYukawa}
 respectively.  In section \ref{Section: Doublets} we will present the doublet model starting with the specifics and continuing with the RGEs.  In Section \ref{Section: Triplets} we will follow a similar format for the triplet model.

\section{Doublet Model}
%
\label{Section: Doublets}
%
\subsection{Particle Content}
%
\label{Section: Doublets: Particle Content}
$\indent$ In Table \ref{Tbl: Particle Content Doublets} is the particle content for the doublet implementation of SUSYLR and the particle representations under the non-abelian gauge groups.  The particle quantum numbers are stated for the $U(1)_{B-L}$ gauge group ({The $B - L$ number used in the RGEs follows the GUT normalization scheme; the values in the table do not.  To get the GUT-normalized value, multiply the number in the table by $\sqrt{3/8}$}).
\begin{table}[!h]
\begin{center}
\begin{tabular}{lccccccc}
		& $SU(3)^c$	&$\times$& $SU(2)_L$	&$\times$& $SU(2)_R$	&$\times$& $U(1)_{B-L}$		\\
$Q$		& 3		&	 & 2		&	 & 1		&	 & $+\third$		\\
$Q^c$		& 3		&	 & 1		&	 & 2		&	 & $-\third$		\\
$L$		& 1		&	 & 2		&	 & 1		&	 & $-1$			\\
$L^c$		& 1		&	 & 1		&	 & 2		&	 & $+1$			\\
$\Phi_a$	& 1		&	 & 2		&	 & 2		&	 & $0$			\\
$\chi$		& 1		&	 & 2		&	 & 1		&	 & $+1$			\\
$\chi^c$	& 1		&	 & 1		&	 & 2		&	 & $-1$			\\
$\bar{\chi}$	& 1		&	 & 2		&	 & 1		&	 & $-1$			\\
$\bar{\chi}^c$	& 1		&	 & 1		&	 & 2		&	 & $+1$			\\
$S^\alpha$	& 1		&	 & 1		&	 & 1		& 	 & $0$
\end{tabular}
\caption{This table shows the representations for the non-abelian gauge groups and the $B-L$ number for $U(1)$}
\label{Tbl: Particle Content Doublets}
\end{center}
\end{table}

The $Q$ and $L$ are the quark and lepton fields of the MSSM and Q\up{c} and L\up{c} are the equivalent 
$SU(2)_R$ fields.  In order to keep this model general, we allow for an arbitrary amount of singlet fields, $n_S$  and so in $S^\alpha$, $\alpha = 1 \ldots n_S$.  Likewise there are $n_\Phi$ bidoublet fields and so in $\Phi_a$, $a = 1 \ldots n_\Phi$.  
Note that while including one $\Phi$ bidoublet does give mass to the fermions, it does not produce quark mixings at tree level (thus another method is required for $V_{CKM} \ne 1$ -- see, for instance \cite{ArXiv:KSBabu.BDutta.RNMohapatra:PartialYukawa}) and so most models set $n_\Phi = 2$.
  The SU(2) doublets and bidoublets are represented in the following manner (with color and generational indices suppressed):
$$
\begin{array}{cc}
\displaystyle
Q = \colTwoVect{u}{d}
& \displaystyle
Q^c = \colTwoVect{d^c}{-u^c}
\\
\multicolumn{2}{c}{ \displaystyle
\Phi_a = \twoByTwo	{\nop{\phi_d^0}_a}		{\nop{\phi_u^+}_{\!\! a}}
			{\nop{\phi_d^-}_{\!\! a}}	{\nop{\phi_u^0}_a}
		  }
\end{array}
$$
Here $Q$ and $Q^c$ are used as an example for any SU(2) doublet pair.  The other doublets can be written in a 
similar fashion where the charges of the fields must obey the equation $Q = I_{L3} + I_{R3} + \frac{B-L}{2}$
Where $I_{L 3}$ and $I_{R 3}$ are the third component of the $SU(2)_{L}$ and $SU(2)_{R}$
 quantum numbers.
 
Under $SU(2)$, these fields transform as:
$$
\begin{array}{ccccccc}
Q \rightarrow U_L Q						& &
Q^c \rightarrow U_R Q^c						\\ 
L \rightarrow U_L L						& & 
L^c \rightarrow U_R L^c						\\
\chi \rightarrow U_L \chi					& & 
\chi^c \rightarrow U_R \chi^c					\\ 
\bar{\chi} \rightarrow U_L \bar{\chi}				& & 
\bar{\chi}^c \rightarrow U_R \bar{\chi}^c			\\
\Phi_a  \rightarrow U_L \Phi_a U_R^\dagger			& &
S^{\alpha} \rightarrow S^{\alpha}
\end{array}
$$
\\
And their parity transformations are:
%
%
%
$$
\begin{array}{ccccccc}
Q \rightarrow -i \tau_2 Q^{c \, *}				& & 
Q^c \rightarrow i \tau_2 \conj{Q}				\\ 
L \rightarrow -i \tau_2 L^{c \, *}				& & 
L^c \rightarrow i \tau_2 \conj{L}				\\
\chi \rightarrow -i \tau_2 \chi^{c \, *}			& & 
\chi^c \rightarrow i \tau_2 \conj{\chi}				\\ 
\bar{\chi} \rightarrow -i \tau_2 \bar{\chi}^{c \, *}		& & 
\bar{\chi}^c \rightarrow i \tau_2 \conj{\bar{\chi}}		\\
\Phi_a  \rightarrow \Phi_a^\dagger				& &
S^{\alpha} \rightarrow S^{\alpha \, *}
\end{array}
$$
%

%
\subsection{Superpotential and Soft Breaking Lagrangian}

The most general superpotential and soft supersymmetry breaking lagrangian for this model are:
\begin{eqnarray}
W	& =	&   i h_a Q^T \tau_2 \Phi_a Q^c
		  + i h_a^\prime L^T \tau_2 \Phi_a L^c
		  + i \lambda_a \chi^T \tau_2 \Phi_a \chi^c
		  + i \bar{\lambda_a} \bar{\chi}^T \tau_2 \Phi_a \bar{\chi}^c
\nonumber	\\ 
	& 	& + \; i \mu_{\chi}^\alpha S^\alpha \chi^T \tau_2 \bar{\chi} 
		  + i \mu_{\chi^c}^\alpha S^\alpha \chi^{c T} \tau_2 \bar{\chi}^c
		  + \sixth Y^{\alpha\beta\gamma} S^\alpha S^\beta S^\gamma
		  + \mu^\alpha_{ab} S^\alpha \Tr \inp{\Phi_a^{T} \tau_2 \Phi_b \tau_2}
\nonumber	\\ 
	&	& + \; i M_{\chi} \chi^T \tau_2 \bar{\chi} + i M_{\chi^c} \chi^{c T} \tau_2 \bar{\chi}^c
		  + M_{ab} \Tr \inp{\Phi_a^{T} \tau_2 \Phi_b \tau_2}
\nonumber	\\ 
	&	& + \; \half M_S^{\alpha \beta} S^\alpha S^\beta + L^\alpha S^\alpha
		+ W_{NR}
\end{eqnarray}

\begin{eqnarray}
	\cal{L}_{SB}
	& =	& -\half \inp
			{
				  M_3 \superp{g} \superp{g} 
				+ M_L \superp{W_L} \superp{W_L}
				+ M_R \superp{W_R} \superp{W_R}
				+ M_1 \superp{B} \superp{B}
				+ \mbox{h.c.}
			    }
 \nonumber	\\
	&	& - \; \inb
			{
				  i A_{Q a} \superp{Q}^T \tau_2 \Phi_a \superp{Q^c}
				+ i A_{L a} \superp{L}^T \tau_2 \Phi_a \superp{L^c}
				+ i A_{\lambda a} \chi \tau_2 \Phi_a \chi^c
			 	+ i A_{\bar\lambda a} \bar\chi \tau_2 \Phi_a \bar\chi^c
			\right.
 \nonumber	\\
	& 	&
			\indent
			\left.
		  		+ \; i A_\chi^{\alpha} S^\alpha \chi \tau_2 \bar\chi
		  		+ i A_{\chi^c}^{\alpha} S^\alpha \chi^c \tau_2 \bar\chi^c
		  		+ \sixth A_S^{\alpha\beta\gamma} S^\alpha S^\beta S^\gamma
				+ A_{\Phi a b}^{\alpha} S^\alpha \Tr \inp
				{
					\Phi_a^{T} \tau_2 \Phi_b \tau_2
				}
				+ \mbox{h.c.}
			}
 \nonumber	\\
	&	& - \inb
		{
			  i B_\chi \chi \tau_2 \bar\chi + i B_{\chi^c} \chi^c \tau_2 \bar\chi^c
			+ B_{ab} \Tr \inp
			{
				\Phi_a^{T} \tau_2 \Phi_b \tau_2 
			}
			+ \half B_s^{\alpha\beta} S^\alpha S^\beta
		}
 \nonumber	\\
	&	& - \left[
			  m_Q^2 \superp{Q}^T \superp{Q}^*
			+ m_{Q^c} \superp{Q^c}^\dagger \superp{Q^c}
			+ m_{L}^2 \superp{L}^T \superp{L}^*
		  	+ m_{L^c}^2 \superp{L^c}^\dagger \superp{L^c}
		  	+ m_{\chi}^2 \chi^\dagger \chi + m_{\chi^c}^2 \chi^{c\dagger} \chi^c
		\right.
\nonumber	\\
 	&	&
 		\indent
 		\left.
		 	+ \; m_{\bar\chi}^2 \bar{\chi}^\dagger \bar\chi
			+ m_{\bar\chi^c}^2 \bar{\chi}^{c \dagger} \bar\chi^c
			+ m_{\Phi ab}^2 \Tr \inp{ \Phi_a^\dagger \Phi_b }
			+ \inp{m_S^2}^{\alpha \beta} {\conj{S}}^\alpha S^\beta
		\right]
\end{eqnarray}
Where we have suppressed the generational and SU(2) indices.  If these were to be included, the term $m_Q^2 \superp{Q}^T \superp{Q}^*$ would be written as $\inp{m_Q^2}^j_i \superp{Q}_{j \alpha} \superp{Q}_\alpha^i $, where the lower case english letters run over generations and the greek letters run over $SU(2)_L$ index in this case.  $W_{NR}$ denotes non-renormalizable terms arising from higher scale physics and would include 
$(f L \tau_2 L \chi \tau_2 \chi + f^* L^c \tau_2 L^c \chi^c \tau_2 \chi^c)/M_{pl}$
the term that gives rise to the seasaw mechanism, ($M_{pl}$ refers to the higher scale physics, e.g. the planck scale).  Since coefficients of this form are suppressed by the scale of higher physics, their contributions to the renormalization group equations may be ignored and will not be included below.

By demanding parity invariance from this theory, we also find the following relations among the parameters \cite{Journal:Phys.Rev.Lett.76:3490.1996, Journal:Phys.Rev.Lett.76:3486.1996, Journal:Phys.Rev.Lett.79:4744.1997}: yuakawa couplings are hermitian except for $\mu_\chi^\alpha$ and $\mu_{\chi^c}^\alpha$, trilinear couplings are hermitian except for $A_\chi^\alpha$ and $A_{\chi^c}^\alpha$, soft breaking mass terms for an $SU(2)_L$ doublet are equal to those of the corresponding $SU(2)_R$ doublet and 
$$
\begin{array}{*{13}{c}}
\multicolumn{13}{c}{
	\begin{array}{*{11}{c}}
		\mu_\chi^\alpha = \mu_{\chi^c}^{\alpha *}	&	&
		A_\chi^\alpha = A_{\chi^c}^{\alpha *}		&	&
		M_\chi = M_{\chi^c}^*				&	&
		M_{a b} = M_{a b}^*				&	&
		M_S^{\alpha \beta} = M_S^{\alpha \beta *}	&	&
		L^\alpha = L^{\alpha *}
	\end{array}
}
\\
g_L = g_R						&	&
M_1 = M_1^*						&	&
M_L = M_R^*						&	&
M_3 = M_3^*						&	&
B_\chi = B_{\chi^c}^*					&	&
B_{\Phi a b} = B_{\Phi a b}^*				&	&
B_S^{\alpha \beta} = B_S^{\alpha \beta *}		
\end{array}
$$
Where $g_L$ and $g_R$ are the $SU(2)_L$ and $SU(2)_R$ gauge coupling constants, respectively.
\subsection{RGEs}
$\indent$ In this section we present our results: one loop renormalization group equations for this model.  The equations are broken up into subsections corresponding to their coupling type.
\subsubsection{Gauge Couplings}
%
\vspace{-.6cm}
\begin{equation}
\begin{array}{*{3}{c}}
\displaystyle
16\pi^2 \deriv{}{t} g_1 = 9 g_1^{3}				&	& \displaystyle
16\pi^2 \deriv{}{t} g_L	= (1 + n_\Phi) g_L^{3}			\\	 \displaystyle
16\pi^2 \deriv{}{t} g_R = (1 + n_\Phi) g_R^{3}			&	& \displaystyle
16\pi^2 \deriv{}{t} g_3 = -3 g_3^{3}
\end{array}
\end{equation}


\subsubsection{Yukawa Couplings}
%
\vspace{-.5cm}
\begin{eqnarray}
16\pi^2 \deriv{}{t} h_a
	& =	&  h_a \left[
			  2 h_b^\dagger h_b 
		   	- \sixth g_1^2
			- 3 g_L^2
			- 3 g_R^2
			- \frac{16}{3} g_3^2
			\right]
 \nonumber	\\
	&	& + \; h_b \left[ 
			  \Tr\inp{3 h_b^\dagger h_a + h_b^{\prime\dagger} h_a^\prime}
			+ 2 h_b^\dagger h_a
			+ \conj{\lambda}_b \lambda_a + \conj{\bar{\lambda}}_b \bar{\lambda}_a
			+ 4\inp{\mu_\Phi^{\alpha \dagger} \mu_\Phi^\alpha}_{ba}
			\right]
	\\
 \nonumber	\\
16\pi^2 \deriv{}{t} h_a^{\prime}
	& =	&  h_a^{\prime} \left[
			  2 h_b^{\prime \dagger} h_b^{\prime}
		   	- \frac{3}{2} g_1^2
			- 3 g_L^2 
			- 3 g_R^2
			\right]
 \nonumber	\\
	&	& + \; h_b^{\prime} \left[ 
			  \Tr\inp{3 h_b^\dagger h_a + h_b^{\prime\dagger} h_a^\prime}
			+ 2 h_b^{\prime \dagger} h_a^\prime
			+ \conj{\lambda}_b \lambda_a + \conj{\bar{\lambda}}_b \bar{\lambda}_a
			+ 4\inp{\mu_\Phi^{\alpha \dagger} \mu_\Phi^\alpha}_{ba}
			\right]
	\\
 \nonumber	\\
16\pi^2 \deriv{}{t} \lambda_a
	& =	&	\lambda_a
			\inb
			{
				\mu_{\chi}^{\alpha*} \mu_{\chi}^{\alpha}
				+ \mu_{\chi^c}^{\alpha*} \mu_{\chi^c}^{\alpha} 
				+ 4 \conj\lambda_b \lambda_b - \frac{3}{2} g_1^2 - 3 g_L^2 - 3 g_R^2
			}
 \nonumber	\\
	&	&	+ \; \lambda_b
			\inb
			{
				  \Tr\inp{3 h_b^\dagger h_a + h_b^{\prime\dagger} h_a^\prime}
				+ \conj{\lambda}_b \lambda_a + \conj{\bar{\lambda}}_b \bar{\lambda}_a
				+ 4\inp{\mu_\Phi^{\alpha \dagger	} \mu_\Phi^\alpha}_{ba}
			}
	\\
 \nonumber	\\
16\pi^2 \deriv{}{t} \bar{\lambda}_a
	& =	&	\bar{\lambda}_a
			\inb
			{
				  \mu_{\chi}^{\alpha*} \mu_{\chi}^{\alpha}
				+ \mu_{\chi^c}^{\alpha*} \mu_{\chi^c}^{\alpha}
				+ 4 \conj{\bar\lambda}_b \bar\lambda_b - \frac{3}{2} g_1^2 - 3 g_L^2 - 3 g_R^2 
			}
 \nonumber	\\
	&	&	+ \; \bar\lambda_b
			\inb
			{
				  \Tr\inp{3 h_b^\dagger h_a + h_b^{\prime\dagger} h_a^\prime}
				+ \conj{\lambda}_b \lambda_a + \conj{\bar{\lambda}}_b \bar{\lambda}_a
				+ 4\inp{\mu_\Phi^{\alpha \dagger} \mu_\Phi^\alpha}_{ba}
			}
	\\
 \nonumber	\\
16\pi^2 \deriv{}{t} \mu_{\chi}^{\alpha}
	& =	&	\mu_{\chi}^{\alpha}
			\inb
			{
				  2 \conj\lambda_a \lambda_a + 2 \conj{\bar\lambda}_a \bar\lambda_a
				+ 2 \mu_{\chi}^{\beta*} \mu_{\chi}^{\beta}
				 - \frac{3}{2} g_1^2 - 3 g_L^2
			}
 \nonumber	\\
	&	&	+ \; \mu_{\chi}^{\beta}
			\inb
			{
				  2 \mu_{\chi}^{\beta*} \mu_\chi^{\alpha}
				+ 2 \mu_{\chi^c}^{\beta*} \mu_{\chi^c}^{\alpha}
				+ \half \inp{Y^{\beta\gamma\delta}}^{*} Y^{\alpha\gamma\delta}
				+ 8\Tr\inp
					{
						\mu_\Phi^{\beta \dagger} \mu_\Phi^{\alpha}
					} 
			}
	\\
 \nonumber	\\
 \nonumber	\\ 
 \nonumber	\\
16\pi^2 \deriv{}{t} \mu_{\chi^c}^{\alpha}
	& =	&	\mu_{\chi^c}^{\alpha}
			\inb
			{
				 2 \conj\lambda_a \lambda_a +  2 \conj{\bar\lambda}_a \bar\lambda_a
				+ 2 \mu_{\chi^c}^{\beta*} \mu_{\chi^c}^{\beta}
				 - \frac{3}{2} g_1^2 - 3 g_R^2
			}
 \nonumber	\\
	&	&	+ \; \mu_{\chi^c}^{\beta}
			\inb
			{
				  2 \mu_{\chi}^{\beta*} \mu_\chi^{\alpha}
				+ 2 \mu_{\chi^c}^{\beta*} \mu_{\chi^c}^{\alpha} 
				+ \half \conj{\inp{Y^{\beta\mu\nu}}} Y^{\alpha\mu\nu}
				+ 8\Tr\inp
					{
						\mu_\Phi^{\beta \dagger} \mu_\Phi^{\alpha}
					} 
			}
	\\
 \nonumber	\\
16\pi^2 \deriv{}{t} Y^{\alpha\beta\gamma}
	& =	&	Y^{\alpha \beta \rho}
			\inb
			{
				  2 \mu_{\chi}^{\rho *} \mu_\chi^{\gamma}
				+ 2 \mu_{\chi^c}^{\rho *} \mu_{\chi^c}^{\gamma} 
				+ \half \conj{\inp{Y^{\rho \mu \nu}}} Y^{\gamma \mu \nu}
				+ 8\Tr\inp
					{
						\mu_\Phi^{\rho \dagger} \mu_\Phi^{\gamma}
					}
			}
 \nonumber	\\
	& 	&	+ \; Y^{\gamma \beta \rho}
			\inb
			{
				  2 \mu_{\chi}^{\rho *} \mu_\chi^{\alpha}
				+ 2 \mu_{\chi^c}^{\rho *} \mu_{\chi^c}^{\alpha} 
				+ \half \conj{\inp{Y^{\rho \mu \nu}}} Y^{\alpha \mu \nu}
				+ 8\Tr\inp
					{
						\mu_\Phi^{\rho \dagger} \mu_\Phi^{\alpha}
					}
			}
 \nonumber	\\
	& 	&	+ \; Y^{\gamma\alpha\rho}
			\inb
			{
				  2 \mu_{\chi}^{\rho*} \mu_\chi^{\beta} 
				+ 2 \mu_{\chi^c}^{\rho*} \mu_{\chi^c}^{\beta}
				+ \half \conj{\inp{Y^{\rho \mu \nu}}} Y^{\beta \mu \nu}
				+ 8\Tr\inp
					{
						\mu_\Phi^{\rho \dagger} \mu_\Phi^{\beta}
					}
			}
	\\
 \nonumber	\\
16\pi^2 \deriv{}{t} \mu_{\Phi a b}^{\alpha}
	& =	&	\mu_{\Phi a c}^{\alpha}
			\inb
			{
				  \Tr\inp{3 h_c^\dagger h_b + h_c^{\prime\dagger} h_b^\prime}
				+ \conj{\lambda}_c \lambda_b + \conj{\bar{\lambda}}_c \bar{\lambda}_b
				+ 4\inp{\mu_\Phi^{\beta \dagger} \mu_\Phi^\beta}_{cb}
			}
 \nonumber	\\
	& 	&	+ \inb
			{
				  \Tr\inp{3 h_a h_c^\dagger + h_a^\prime h_c^{\prime\dagger}}
				+ \lambda_a \conj{\lambda}_c + \bar{\lambda}_a \conj{\bar{\lambda}}_c 
				+ 4\inp{\mu_\Phi^{\beta} \mu_\Phi^{\beta \dagger}}_{ac}
			}
			\mu_{\Phi c b}^{\alpha}
\nonumber	\\
	& 	&	+ \; \mu_{\Phi a b}^{\beta}
			\inb
			{
				   2 \mu_{\chi}^{\beta*} \mu_\chi^{\alpha}
				+ 2 \mu_{\chi^c}^{\beta*} \mu_{\chi^c}^{\alpha} 
				+ \half \conj{\inp{Y^{\beta \mu \nu}}} Y^{\alpha \mu \nu}
				+ 8\Tr\inp
					{
						\mu_\Phi^{\beta \dagger} \mu_\Phi^{\alpha}
					}
			}
\nonumber	\\
	&	&
			- \mu_{\Phi a b}^{\alpha} \inp{ 3 g_L^{2} + 3 g_R^{2} }
\end{eqnarray}

\subsubsection{Mass Couplings}
%
\vspace{-.5cm}
\begin{eqnarray}
16\pi^2 \deriv{}{t} M_{\chi}
	& =	&  	M_{\chi}
			\inb
			{
				2 \conj\lambda_a \lambda_a + 2 \conj{\bar\lambda}_a \bar\lambda_a
				+ 2 \mu_{\chi}^{\alpha*} \mu_{\chi}^{\alpha}
				 - \frac{3}{2} g_1^2 - 3 g_L^{2}
			}
	\\
 \nonumber	\\
16\pi^2 \deriv{}{t} M_{\chi^c}
	& =	&  	M_{\chi^c}
			\inb
			{
				2 \conj\lambda_a \lambda_a + 2 \conj{\bar\lambda}_a \bar\lambda_a
				+ 2 \mu_{\chi^c}^{\alpha*} \mu_{\chi^c}^{\alpha}
				 - \frac{3}{2} g_1^2 - 3 g_R^{2}
			}
	\\
 \nonumber	\\
16\pi^2 \deriv{}{t} M_{\Phi a b}
	& =	&  	M_{\Phi a c}
			\inb
			{
				  \Tr\inp
					{
						3 h_c^\dagger h_b + h_c^{\prime\dagger} h_b^\prime
					}
				+ \conj{\lambda}_c \lambda_b + \conj{\bar{\lambda}}_c \bar{\lambda}_b
				+ 4\inp{\mu_\Phi^{\alpha \dagger} \mu_\Phi^\alpha}_{cb}
			}
 \nonumber	\\
	& 	&
			+ \;
			\inb
			{
				  \Tr\inp
					{
						3 h_a h_c^\dagger + h_a^\prime h_c^{\prime\dagger}
					}
				+ \lambda_a \conj{\lambda}_c + \bar{\lambda}_a \conj{\bar{\lambda}}_c 
				+ 4\inp{\mu_\Phi^{\alpha} \mu_\Phi^{\alpha \dagger}}_{ac}
			}
			M_{\Phi c b}
 \nonumber	\\
	&	&	+ \; M_{\Phi a b} \inp
			{
				- 3 g_L^2 - 3 g_R^2
			}
	\\
 \nonumber	\\
16\pi^2 \deriv{}{t} M_s^{\alpha\beta}
	& =	&	M_s^{\alpha\rho}
			\inb
			{
				  2 \mu_{\chi^c}^{\rho*} \mu_{\chi^c}^{\beta} 
				+ 2 \mu_{\chi}^{\rho*} \mu_\chi^{\beta}
				+ \half \conj{\inp{Y^{\rho \mu \nu}}} Y^{\beta \mu \nu}
				+ 8\Tr\inp
					{
						\mu_\Phi^{\rho \dagger} \mu_\Phi^{\beta}
					}
			}
\nonumber	\\
	& 	&	+ \; M_s^{\beta \rho}
			\inb
			{
				  2 \mu_{\chi^c}^{\rho *} \mu_{\chi^c}^{\alpha} 
				+ 2 \mu_{\chi}^{\rho *} \mu_\chi^{\alpha}
				+ \half \conj{\inp{Y^{\rho \mu \nu}}} Y^{\alpha \mu \nu}
				+ 8\Tr\inp
					{
						\mu_\Phi^{\rho \dagger} \mu_\Phi^{\alpha}
					}
			}
\end{eqnarray}

\subsubsection{Linear Term}
%
\vspace{-.5cm}
\begin{eqnarray}
16\pi^2 \deriv{}{t} L^{\alpha}
	& =	&	L^{\beta}
			\inb
			{
				  2 \mu_{\chi^c}^{\beta*} \mu_{\chi^c}^{\alpha} 
				+ 2 \mu_{\chi}^{\beta*} \mu_\chi^{\alpha}
				+ \half \inp{Y^{\beta \mu \nu}}^{*} Y^{\alpha \mu \nu}
				+ 8\Tr\inp
					{
						\mu_\Phi^{\beta \dagger} \mu_\Phi^{\alpha}
					}
			}
\end{eqnarray}

\subsubsection{Gaugino Masses}
%
\vspace{-.5cm}
\begin{equation}
\begin{array}{*{7}{c}}
\displaystyle
	16\pi^2 \deriv{}{t} M_1	= 18 M_1 g_1^{2}				&	& \displaystyle
	16\pi^2 \deriv{}{t} g_L = 2(1 + n_\Phi) M_L g_L^{2}			\\ \nonumber \displaystyle
	16\pi^2 \deriv{}{t} g_R = 2(1 + n_\Phi) M_R g_R^{2}			&	& \displaystyle
	16\pi^2 \deriv{}{t} g_3 = -6 M_3 g_3^{2}
\end{array}
\end{equation}

\subsubsection{Soft Breaking Trilinear $A$'s}
%
\vspace{-.5cm}
\begin{eqnarray}
	16 \pi^2 \deriv{}{t} A_{Q a}
		& =	& A_{Q a} \inb
				{
					  2 h_b^\dagger h_b
					- \frac{1}{6} g_1^2
					- 3 g_L^2 
					- 3 g_R^2
					- \frac{16}{3} g_3^2
			      }
 \nonumber	\\
	&	&
		  + \; h_a \inb
				{	  
					  4 h_b^\dagger A_{Q b}
		  			+ \frac{1}{3} M_1 g_1^2
		  			+ 6 M_L g_L^2
		  			+ 6 M_R g_R^2
		  			+ \frac{32}{3} M_3 g_3^2
		  		}
 \nonumber	\\
	&	&
		  + \; h_{b} \inb
				{
					2 h_b^{\dagger} A_{Qa} + \Tr \inp
					{
						6 h_b^{\dagger} A_{Qa} + 2 h_b^{\prime\dagger} A_{La}
					}
					+ 2 \lambda_b^{*} A_{\lambda a}
					+ 2 \bar{\lambda}_b^{*} A_{\bar\lambda a}
					+ 8 \inp
						{
							\mu_\Phi^{\alpha \dagger} A_\Phi^{\alpha}
						}_{ba}
				}
 \nonumber	\\
	&	&
		  + \; A_{Q b} \inb
				{ 
					4 h_b^{\dagger} h_a + \Tr\inp
					{
						3 h_b^\dagger h_a + h_b^{\prime\dagger} h_a^{\prime}
					}
					+ \lambda_b^{*} \lambda_a 
					+ \bar\lambda_b^{*} \bar\lambda_a
					+ 4 \inp
					{
						\mu_\Phi^{\alpha \dagger} \mu_\Phi^\alpha
					}_{ba}
				}
	\\
 \nonumber	\\
	16 \pi^2 \deriv{}{t} A_{L a}
		& =	& A_{L a} \inb
				{
					  2 h_b^{\prime \dagger} h_b^\prime
					- \frac{3}{2} g_1^2
					- 3 g_L^2 
					- 3 g_R^2
			      }
 \nonumber	\\
	&	&
		  + \; h_a^{\prime} \inb
				{	  
					  4 h_b^{\prime \dagger} A_{L b}
		  			+ 3 M_1 g_1^2
		  			+ 6 M_L g_L^2
		  			+ 6 M_R g_R^2
		  		}
 \nonumber	\\
	&	&
		  + \; h_{b}^{\prime} \inb
				{
					2 h_b^{\prime \dagger} A_{La} + \Tr \inp
					{
						6 h_b^{\dagger} A_{Qa} + 2 h_b^{\prime\dagger} A_{La}
					}
					+ 2 \lambda_b^{*} A_{\lambda a}
					+ 2 \bar{\lambda}_b^{*} A_{\bar\lambda a}
					+ 8 \inp
						{
							\mu_\Phi^{\alpha \dagger} A_\Phi^{\alpha}
						}_{ba}
				}
 \nonumber	\\
	&	&
		  + \; A_{L b} \inb
				{ 
					4 h_b^{\prime \dagger} h_a^{\prime} + \Tr\inp
					{
						3 h_b^\dagger h_a + h_b^{\prime\dagger} h_a^{\prime}
					}	
					+ \lambda_b^{*} \lambda_a 
					+ \bar\lambda_b^{*} \bar\lambda_a
					+ 4 \inp
					{
						\mu_\Phi^{\alpha \dagger} \mu_\Phi^\alpha
					}_{ba}
				}
	\\
 \nonumber	\\
%
%
	16 \pi^2 \deriv{}{t} A_{\lambda a}
		& =	& A_{\lambda a} \inb
				{
					  4 \lambda_b^{*} \lambda_b
					+ \mu_{\chi}^{\alpha *} \mu_{\chi}^\alpha
					+ \mu_{\chi^c}^{\alpha *} \mu_{\chi^c}^\alpha
					- \frac{3}{2} g_1^2
					- 3 g_L^2 
					- 3 g_R^2
			      }
 \nonumber	\\
	&	&
		  + \; \lambda_a \inb
				{	  
					  8 \lambda_b^{*} A_{\lambda b}
					+ 2 \mu_{\chi}^{\alpha *} A_{\chi}^{\alpha}
					+ 2 \mu_{\chi^c}^{\alpha *} A_{\chi^c}^{\alpha}
		  			+ 3 M_1 g_1^2
		  			+ 6 M_L g_L^2
		  			+ 6 M_R g_R^2
		  		}
 \nonumber	\\
	&	&
		  + \; A_{\lambda b} \inb
				{ 
					\Tr\inp
					{
						3 h_b^\dagger h_a + h_b^{\prime\dagger} h_a^{\prime}
					}
					+ \lambda_b^{*} \lambda_a 
					+ \bar\lambda_b^{*} \bar\lambda_a
					+ 4 \inp
					{
						\mu_\Phi^{\alpha \dagger} \mu_\Phi^\alpha
					}_{ba}
				}
 \nonumber	\\
	&	&
		  + \; \lambda_{b} \inb
				{
					\Tr \inp
					{
						6 h_b^{\dagger} A_{Qa} + 2 h_b^{\prime\dagger} A_{La}
					}
					+ 2 \lambda_b^{*} A_{\lambda a}
					+ 2 \bar{\lambda}_b^{*} A_{\bar\lambda a}
					+ 8 \inp
						{
							\mu_\Phi^{\alpha \dagger} A_\Phi^{\alpha}
						}_{ba}
				}
	\\
 \nonumber	\\
	16 \pi^2 \deriv{}{t} A_{\bar\lambda a}
		& =	& A_{\bar\lambda a} \inb
				{
					  4 \bar\lambda_b^{*} \bar\lambda_b
					+ \mu_{\chi}^{\alpha *} \mu_{\chi}^\alpha
					+ \mu_{\chi^c}^{\alpha *} \mu_{\chi^c}^\alpha
					- \frac{3}{2} g_1^2
					- 3 g_L^2 
					- 3 g_R^2
			      }
 \nonumber	\\
	&	&
		  + \; \bar\lambda_a \inb
				{	  
					  8 \bar\lambda_b^{*} A_{\bar\lambda b}
					+ 2 \mu_{\chi}^{\alpha *} A_{\chi}^{\alpha}
					+ 2 \mu_{\chi^c}^{\alpha *} A_{\chi^c}^{\alpha}
		  			+ 3 M_1 g_1^2
		  			+ 6 M_L g_L^2
		  			+ 6 M_R g_R^2
		  		}
 \nonumber	\\
	&	&
		  + \; A_{\bar\lambda b} \inb
				{ 
					\Tr\inp
					{
						3 h_b^\dagger h_a + h_b^{\prime\dagger} h_a^{\prime}
					}
					+ \lambda_b^{*} \lambda_a 
					+ \bar\lambda_b^{*} \bar\lambda_a
					+ 4 \inp
					{
						\mu_\Phi^{\alpha \dagger} \mu_\Phi^\alpha
					}_{ba}
				}
 \nonumber	\\
	&	&
		  + \; \bar\lambda_{b} \inb
				{
					\Tr \inp
					{
						6 h_b^{\dagger} A_{Qa} + 2 h_b^{\prime\dagger} A_{La}
					}
					+ 2 \lambda_b^{*} A_{\lambda a}
					+ 2 \bar{\lambda}_b^{*} A_{\bar\lambda a}
					+ 8 \inp
						{
							\mu_\Phi^{\alpha \dagger} A_\Phi^{\alpha}
						}_{ba}
				}
	\\
 \nonumber	\\
16\pi^2 \deriv{}{t} A_{\chi}^{\alpha}
	& =	&	A_{\chi}^{\alpha}
			\inb
			{
				   2 \conj\lambda_a \lambda_a + 2 \conj{\bar\lambda}_a \bar\lambda_a
				+ 2 \mu_{\chi}^{\beta*} \mu_{\chi}^{\beta}
				 - \frac{3}{2} g_1^2 - 3 g_L^2
			}
 \nonumber	\\
	&	&	+ \; \mu_{\chi}^{\alpha}
			\inb
			{
				   4 \conj\lambda_a A_{\lambda a} + 4 \conj{\bar\lambda}_a A_{\bar\lambda a}
				+ 4 \mu_{\chi}^{\beta*} A_{\chi}^{\beta}
				+ 3 M_1 g_1^2 + 6 M_L g_L^2
			}
 \nonumber	\\
	&	&	+ \; A_{\chi^c}^{\beta}
			\inb
			{
				 2 \mu_{\chi}^{\beta*} \mu_\chi^{\alpha}
				+ 2 \mu_{\chi^c}^{\beta*} \mu_{\chi^c}^{\alpha} 
				+ \half \conj{\inp{Y^{\beta\mu\nu}}} Y^{\alpha\mu\nu}
				+ 8\Tr\inp
				{
					\mu_\Phi^{\beta \dagger} \mu_\Phi^{\alpha}
				} 
			}
 \nonumber	\\
	&	&	+ \; \mu_{\chi}^{\beta}
			\inb
			{
				  4 \mu_{\chi}^{\beta*} A_\chi^{\alpha}
				+ 4 \mu_{\chi^c}^{\beta*} A_{\chi^c}^{\alpha} 
				+ \conj{\inp{Y^{\beta\mu\nu}}} A_S^{\alpha\mu\nu}
				+ 16\Tr\inp
					{
						\mu_\Phi^{\beta \dagger} A_\Phi^{\alpha}
					} 
			}
	\\
 \nonumber	\\
16\pi^2 \deriv{}{t} A_{\chi^c}^{\alpha}
	& =	&	A_{\chi^c}^{\alpha}
			\inb
			{
				  2 \conj\lambda_a \lambda_a + 2 \conj{\bar\lambda}_a \bar\lambda_a
				+ 2 \mu_{\chi^c}^{\beta*} \mu_{\chi^c}^{\beta}
				 - \frac{3}{2} g_1^2 - 3 g_R^2
			}
 \nonumber	\\
	&	&	+ \; \mu_{\chi^c}^{\alpha}
			\inb
			{
				  4 \conj\lambda_a A_{\lambda a} + 4 \conj{\bar\lambda}_a A_{\bar\lambda a}
				+ 4 \mu_{\chi^c}^{\beta*} A_{\chi^c}^{\beta}
				+ 3 M_1 g_1^2 + 6 M_R g_R^2
			}
 \nonumber	\\
	&	&	+ \; A_{\chi^c}^{\beta}
			\inb
			{
				  2 \mu_{\chi}^{\beta*} \mu_\chi^{\alpha}
				+ 2 \mu_{\chi^c}^{\beta*} \mu_{\chi^c}^{\alpha} 
				+ \half \conj{\inp{Y^{\beta \mu \nu}}} Y^{\alpha \mu \nu}
				+ 8\Tr\inp
					{
						\mu_\Phi^{\beta \dagger} \mu_\Phi^{\alpha}
					} 
			}
 \nonumber	\\
	&	&	+ \; \mu_{\chi^c}^{\beta}
			\inb
			{
				  4 \mu_{\chi}^{\beta*} A_\chi^{\alpha}
				+ 4 \mu_{\chi^c}^{\beta*} A_{\chi^c}^{\alpha} 
				+ \conj{\inp{Y^{\beta \mu \nu}}} A_S^{\alpha \mu \nu}
				+ 16\Tr\inp
					{
						\mu_\Phi^{\beta \dagger} A_\Phi^{\alpha}
					} 
			}
	\\
 \nonumber	\\
16\pi^2 \deriv{}{t} A_S^{\alpha \beta \gamma}
	& =	&	A_S^{\alpha \beta \rho}
			\inb
			{
				  2 \mu_{\chi}^{\rho*} \mu_\chi^{\gamma}
				+ 2 \mu_{\chi^c}^{\rho*} \mu_{\chi^c}^{\gamma} 
				+ \half \conj{\inp{Y^{\rho \mu \nu}}} Y^{\gamma\mu \nu}
				+ 8\Tr\inp
					{
						\mu_\Phi^{\rho \dagger} \mu_\Phi^{\gamma}
					}
			}
 \nonumber	\\
	&	&	+ \; Y^{\alpha\beta\rho}
			\inb
			{
				  4 \mu_{\chi}^{\rho*} A_\chi^{\gamma}
				+ 4 \mu_{\chi^c}^{\rho*} A_{\chi^c}^{\gamma}
				+ \conj{\inp{Y^{\rho\mu \nu}}} A_S^{\gamma\mu \nu}
				+ 16\Tr\inp
					{
						\mu_\Phi^{\rho \dagger} A_\Phi^{\gamma}
					}
			}
 \nonumber	\\
	& 	&	+ \; A_S^{\gamma\beta\rho}
			\inb
			{
				  2 \mu_{\chi}^{\rho*} \mu_\chi^{\alpha}
				+ 2 \mu_{\chi^c}^{\rho*} \mu_{\chi^c}^{\alpha}
				+ \half \conj{\inp{Y^{\rho\mu \nu}}} Y^{\alpha\mu \nu}
				+ 8\Tr\inp
					{
						\mu_\Phi^{\rho \dagger} \mu_\Phi^{\alpha}
					}
			}
 \nonumber	\\
	& 	&	+ \; Y^{\gamma\beta\rho}
			\inb
			{
				  4 \mu_{\chi}^{\rho*} A_\chi^{\alpha}
				+ 4 \mu_{\chi^c}^{\rho*} A_{\chi^c}^{\alpha} 
				+ \conj{\inp{Y^{\rho\mu \nu}}} A_S^{\alpha\mu \nu}
				+ 16\Tr\inp
					{
						\mu_\Phi^{\rho \dagger} A_\Phi^{\alpha}
					}
			}
 \nonumber	\\
	& 	&	+ \; A_S^{\gamma\alpha\rho}
			\inb
			{
				  2 \mu_{\chi}^{\rho*} \mu_\chi^{\beta}
				+ 2 \mu_{\chi^c}^{\rho*} \mu_{\chi^c}^{\beta} 
				+ \half \conj{\inp{Y^{\rho\mu \nu}}} Y^{\beta\mu \nu}
				+ 8\Tr\inp
					{
						\mu_\Phi^{\rho \dagger} \mu_\Phi^{\beta}
					}
			}
 \nonumber	\\
	& 	&	+ \; Y^{\gamma\alpha\rho}
			\inb
			{
				  4 \mu_{\chi}^{\rho*} A_\chi^{\beta}
				+ 4 \mu_{\chi^c}^{\rho*} A_{\chi^c}^{\beta} 
				+ \conj{\inp{Y^{\rho\mu \nu}}} A_S^{\beta\mu \nu}
				+ 16\Tr\inp
					{
						\mu_\Phi^{\rho \dagger} A_\Phi^{\beta}
					}
			}
	\\	
 \nonumber	\\
16\pi^2 \deriv{}{t} A_{\Phi ab}^{\alpha}
	& =	&	A_{\Phi a c}^{\alpha}
			\inb
			{
				  \Tr\inp{3 h_c^\dagger h_b + h_c^{\prime\dagger} h_b^\prime}
				+ \conj{\lambda}_c \lambda_b + \conj{\bar{\lambda}}_c \bar{\lambda}_b
				+ 4\inp
				{
					\mu_\Phi^{\beta \dagger} \mu_\Phi^\beta
				}_{c b}
			}
 \nonumber	\\
	&	&	+ \; \mu_{\Phi ac}^{\alpha}
			\inb
			{
				  \Tr\inp{6 h_c^\dagger A_{Q b} + 2 h_c^{\prime\dagger} A_{L b}}
				+ 2 \conj{\lambda}_c A_{\lambda b} + 2 \conj{\bar{\lambda}}_c A_{\bar\lambda b}
				+ 8 \inp
				{
					\mu_\Phi^{\beta \dagger} A_\Phi^\beta
				}_{c b}
			}
 \nonumber	\\
	& 	&	+ \inb
			{
				  \Tr\inp{3 h_a h_c^\dagger + h_a^\prime h_c^{\prime\dagger}}
				+ \lambda_a \conj{\lambda}_c + \bar{\lambda}_a \conj{\bar{\lambda}}_c 
				+ 4\inp
				{
					\mu_\Phi^{\beta} \mu_\Phi^{\beta \dagger}
				}_{ac}
			}
			A_{\Phi c b}^{\alpha}
 \nonumber	\\
	& 	&	+ \inb
			{
				  \Tr\inp{6 A_{Q a} h_c^\dagger + 2 A_{L a} h_c^{\prime\dagger}}
				+ 2 A_{\lambda a} \conj{\lambda}_c + 2 A_{\bar\lambda a} \conj{\bar{\lambda}}_c 
				+ 8 \inp
				{
					A_\Phi^{\beta} \mu_\Phi^{\beta \dagger}
				}_{a c}
			}
			\mu_{\Phi c b}^{\alpha}
 \nonumber	\\
	& 	&	+ \; A_{\Phi a b}^{\beta}
			\inb
			{
				  2 \mu_{\chi}^{\beta*} \mu_\chi^{\alpha}
				+ 2 \mu_{\chi^c}^{\beta*} \mu_{\chi^c}^{\alpha} 
				+ \half \conj{\inp{Y^{\beta\mu \nu}}} Y^{\alpha\mu \nu}
				+ 8\Tr\inp
					{
						\mu_\Phi^{\beta \dagger} \mu_\Phi^{\alpha}
					}
			}
 \nonumber	\\
	& 	&	+ \; \mu_{\Phi a b}^{\beta} \!
			\inb
			{
				  4 \mu_{\chi}^{\beta*} A_\chi^{\alpha}
				+ 4 \mu_{\chi^c}^{\beta*} A_{\chi^c}^{\alpha} 
				+ \conj{\inp{Y^{\beta\mu \nu}}} \! A_S^{\alpha\mu \nu}
				+ 16\Tr\inp
				{
					\mu_\Phi^{\beta \dagger} A_\Phi^{\alpha}
				}
			}
\nonumber	\\
	&	&	- \; A_{\Phi a b}^\alpha \inp
			{
				+ 3 g_L^{2} + 3 g_R^{2}
			}
			+ \mu_{\Phi a b}^{\alpha} \inp
			{
				6 M_L g_L^{2} +6 M_R g_R^{2}
			}
\end{eqnarray}

\subsubsection{Soft Breaking Bilinear $B$'s}
%
\vspace{-.5cm}
\begin{eqnarray}
16 \pi^2 \deriv{}{t} B_{\chi}
	& =	& B_{\chi} \inb
		{
			  2 \lambda_a^{*} \lambda_a
			+ 2 \bar\lambda_a^{*} \bar\lambda_a
			+ 2 \mu_{\chi}^{\alpha *} \mu_{\chi}^\alpha
			- \frac{3}{2} g_1^2 
			- 3 g_L^2
		}
 \nonumber	\\
	&	&
		  + \; M_\chi
		\inb
		{	  
			  4 \lambda_a^{*} A_{\lambda a}
			+ 4 \bar\lambda_a^{*} A_{\bar\lambda a}
			+ 4 \mu_{\chi}^{\alpha *} A_{\chi}^{\alpha}
		  	+ 3 M_1 g_1^2
		  	+ 6 M_L g_L^2
		}
 \nonumber	\\
	&	&
		+ \; \mu_\chi^\alpha \inb
		{
			4 \mu_\chi^{\alpha *} B_\chi + 4 \mu_{\chi^c}^{\alpha *} B_{\chi^c}
			+ \conj{\inp{Y^{\alpha \mu \nu }}} B_S^{\mu \nu}
			+16 \Tr \inp
			{
				\mu_\Phi^{\alpha \dagger} B_\Phi
			}
		}
	\\
\nonumber	\\	
16 \pi^2 \deriv{}{t} B_{\chi^c}
	& =	& B_{\chi^c} \inb
		{
			  2 \lambda_a^{*} \lambda_a
			+ 2 \bar\lambda_a^{*} \bar\lambda_a
			+ 2 \mu_{\chi^c}^{\alpha *} \mu_{\chi^c}^\alpha
			- \frac{3}{2} g_1^2 
			- 3 g_R^2
		}
 \nonumber	\\
	&	&
		  + \; M_{\chi^c}
		\inb
		{	  
			  4 \lambda_a^{*} A_{\lambda a}
			+ 4 \bar\lambda_a^{*} A_{\bar\lambda a}
			+ 4 \mu_{\chi^c}^{\alpha *} A_{\chi^c}^{\alpha}
		  	+ 3 M_1 g_1^2
		  	+ 6 M_R g_R^2
		}
 \nonumber	\\
	&	&
		+ \; \mu_{\chi^c}^\alpha \inb
		{
			4 \mu_{\chi}^{\alpha *} B_{\chi} + 4 \mu_{\chi^c}^{\alpha *} B_{\chi^c}
			+ \conj{\inp{Y^{\alpha \mu \nu}}} B_S^{\mu \nu}
			+16 \Tr \inp
			{
				\mu_\Phi^{\alpha \dagger} B_\Phi
			}
		}
	\\
 \nonumber	\\	
16 \pi^2 \deriv{}{t} B_{\Phi a b}
	& =	& B_{\Phi a c} \inb
		{
			\Tr \inp
			{
				3 h_c^\dagger h_b + h_c^{\prime \dagger} h_b ^\prime
			}
			+ \lambda_c^* \lambda_b + \bar\lambda_c^* \bar\lambda_b
			+ 4 \inp
			{
				\mu_\Phi^{\alpha \dagger} \mu_\Phi^\alpha
			}_{c b}
		}
 \nonumber	\\
	&	&
		+ \; M_{\Phi a c} \inb
		{
			  \Tr\inp
			  {
			  	6 h_c^\dagger A_{Q b} + 2 h_c^{\prime\dagger} A_{L b}
			  }
			+ 2 \conj{\lambda}_c A_{\lambda b} + 2 \conj{\bar{\lambda}}_c A_{\bar\lambda b}
			+ 8\inp{\mu_\Phi^{\beta \dagger} A_\Phi^\beta}_{c b}
		}
 \nonumber	\\
	&	&
		+ \; \mu_{\Phi a b}^\alpha \inb
		{
			  4 \mu_{\chi}^{\alpha*} B_\chi
			+ 4 \mu_{\chi^c}^{\alpha*} B_{\chi^c}
			+ \conj{\inp{Y^{\alpha \mu \nu }}} B_S^{\mu \nu}
			+ 16\Tr\inp
			{
				\mu_\Phi^{\alpha \dagger} B_\Phi
			}
		}
 \nonumber	\\
	& 	&
		+ \; \inb
		{
			\Tr \inp
			{
				3 h_a h_c^\dagger + h_a ^\prime h_c^{\prime \dagger}
			}
			+ \lambda_a \lambda_c^* + \bar\lambda_a \bar\lambda_c^*
			+ 4 \inp
			{
				\mu_\Phi^\alpha \mu_\Phi^{\alpha \dagger}
			}_{a c}
		}
		B_{\Phi c b}
 \nonumber	\\
	&	&
		+ \; \inb
		{
			  \Tr\inp
			  {
			  	6 A_{Q a} h_c^\dagger + 2 A_{L a} h_c^{\prime\dagger}
			  }
			+ 2 A_{\lambda a} \conj{\lambda}_c + 2 A_{\bar\lambda a} \conj{\bar{\lambda}}_c
			+ 8\inp{A_\Phi^\alpha \mu_\Phi^{\alpha \dagger}}_{a c}
		}
		M_{\Phi c b}
 \nonumber	\\
	&	&
		- \; B_{\Phi a b} \inp{ 3 g_L^2 + 3 g_R^2}
		+ M_{\Phi a b} \inp{6 M_L g_L^2 + 6 M_R g_R^2}
	\\
 \nonumber	\\
16 \pi^2 \deriv{}{t} B_{S}^{\alpha \beta}
	& =	& 
		B_{S}^{\alpha \rho}
		\inb
		{
			  2 \mu_{\chi}^{\rho *} \mu_\chi^{\beta}
			+ 2 \mu_{\chi^c}^{\rho *} \mu_{\chi^c}^{\beta} 
			+ \half \conj{\inp{Y^{\rho \mu \nu }}} Y^{\beta \mu \nu}
			+ 8\Tr\inp
			{
				\mu_\Phi^{\rho \dagger} \mu_\Phi^{\beta}
			}
		}
 \nonumber	\\
	&	&
		+ \; M_S^{\alpha \rho}
		\inb
		{
			  4 \mu_{\chi}^{\rho *} A_\chi^{\beta}
			+ 4 \mu_{\chi^c}^{\rho *} A_{\chi^c}^{\beta} 
			+ \conj{\inp{Y^{\rho \mu \nu }}} A_S^{\beta \mu \nu}
			+ 16\Tr\inp
			{
				\mu_\Phi^{\rho \dagger} A_\Phi^{\beta}
			}
		}
 \nonumber	\\
	&	&
		+ \; Y^{\alpha \beta \rho}
		\inb
		{
			  4 \mu_{\chi}^{\rho *} B_\chi
			+ 4 \mu_{\chi^c}^{\rho *} B_{\chi^c}
			+ \conj{\inp{Y^{\rho \mu \nu}}} B_S^{\mu \nu}
			+ 16\Tr\inp
			{
				\mu_\Phi^{\rho \dagger} B_\Phi
			}
		}
 \nonumber	\\
	&	&
		+ \; \inb
		{
			  2 \mu_\chi^{\alpha} \mu_{\chi}^{\rho *}
			+ 2 \mu_{\chi^c}^{\alpha} \mu_{\chi^c}^{\rho *}
			+ \half Y^{\alpha \mu \nu} \conj{\inp{Y^{\rho \mu \nu}}}
			+ 8\Tr\inp
			{
				 \mu_\Phi^{\alpha}\mu_\Phi^{\rho \dagger}
			}
		}
		B_S^{\rho \beta}
\nonumber	\\
	&	&
		+ \; \inb
		{
			  4 A_\chi^{\alpha} \mu_{\chi}^{\rho *}
			+ 4 A_{\chi^c}^{\alpha}  \mu_{\chi^c}^{\rho *}
			+ A_S^{\alpha \mu \nu} \conj{\inp{Y^{\rho \mu \nu }}}
			+ 16\Tr\inp
			{
				 A_\Phi^{\alpha} \mu_\Phi^{\rho \dagger}
			}
		}
		M_S^{\rho \beta}
\end{eqnarray}

\subsubsection{Soft Breaking Masses}

For convience, we define the quantity:
\begin{eqnarray}
{\cal S}_2
	& \equiv	&
	4 \inb
	{
		\Tr \inp
		{
		m_Q^2 - m_{Q^c}^2 - m_L^2 + m_{L^c}^2
		}
		+ m_\chi^2 - m_{\chi^c}^2 + m_{\bar\chi^c}^2 - m_{\bar\chi}^2
	}
\end{eqnarray}
Which is used in the soft breaking mass equations below.
\begin{eqnarray}
16 \pi^2 \deriv{}{t} m_Q^{2}
	& =	& 
		  2 m_Q^{2} h_a h_a^{\dagger}
		+ h_a \inp
		{
			2 h_a^{\dagger} m_Q^{2} + 4 m_{Q^c}^{2} h_a^{\dagger} + 4 m_{\Phi a b}^2 h_b^{\dagger}
		}
		 + 4 A_{Q a} A_{Q a}^{\dagger} 
 \nonumber	\\
	&	&
		- \; \third M_1 M_1^{\dagger} g_1^{2} - 6 M_L M_L^{\dagger} g_L^{2} 
		-\frac{32}{3} M_3 M_3^{\dagger} g_3^{2} +\eighth g_1^{2} {\cal S}_2
	\\
 \nonumber	\\
16 \pi^2 \deriv{}{t} m_{Q^c}^2
	& =	&
		  2 m_{Q^c}^2 h_a^{\dagger} h_a
		+ h_a^{\dagger} \inp
		{
			2 h_a m_{Q^c}^2+ 4 m_{Q}^{2} h_a + 4 h_b m_{\Phi b a}^2
		}
		+ 4 A_{Q a}^{\dagger} A_{Q a} 
 \nonumber	\\
	&	&
		- \; \third M_1 M_1^{\dagger} g_1^{2} - 6 M_R M_R^{\dagger} g_R^{2} 
		-\frac{32}{3} M_3 M_3^{\dagger} g_3^{2} - \eighth g_1^{2} {\cal S}_2
	\\
 \nonumber	\\
16 \pi^2 \deriv{}{t} m_{L}^2
	& =	&
		  2 m_L^{2} h_a^{\prime} h_a^{\prime \dagger}
		+ h_a^{\prime} \inp
		{
			2 h_a^{\prime \dagger} m_L^{2} 
			+ 4 m_{L^c}^{2} h_a^{\prime \dagger}
			+ 4 m_{\Phi a b}^2 h_b^{\prime \dagger}
		}
		+ 4 A_{L a} A_{L a}^{\dagger} 
 \nonumber	\\
	&	&
		- \; 3 M_1 M_1^{\dagger} g_1^{2} - 6 M_L M_L^{\dagger} g_L^{2} 
		- \frac{3}{8} g_1^{2} {\cal S}_2
	\\
 \nonumber	\\
16 \pi^2 \deriv{}{t} m_{L^c}^2
	& =	& 2 m_{L^c}^2 h_a^{\prime \dagger} h_a^{\prime}
		+ h_a^{\prime \dagger} \inp
		{
			2 h_a^{\prime} m_{L^c}^2
			+ 4 m_{L}^{2} h_a^{\prime}
			+ 4 h_b^{\prime} m_{\Phi b a}^2
		}
		 + 4 A_{L a}^{\dagger} A_{L a} 
 \nonumber	\\
	&	&
		- \; 3 M_1 M_1^{\dagger} g_1^{2} - 6 M_R M_R^{\dagger} g_R^{2} 
		+ \frac{3}{8} g_1^{2} {\cal S}_2
	\\
 \nonumber	\\
16 \pi^2 \deriv{}{t} m_{\chi}^2
	& =	& 
		  \lambda_a \inb[.5cm]
		{
			  4 m_\chi^{2} \lambda_a^{*}
			+ 4 m_{\chi^c}^{2} \lambda_a^{*}
			+ 4 m_{\Phi a b}^{2} \lambda_b^{*}
		}
		+  \mu_{\chi}^{\alpha} \inb
		{
			2 m_{\chi}^2 \mu_\chi^{\alpha *}
			+ 2 m_{\bar\chi}^2 \mu_\chi^{\alpha *}
			+ 2 \inp{m_{S}^2}^{\alpha \beta} \mu_\chi^{\beta *}
		}
\nonumber	\\
	&	&
		+ \; 4 A_{\lambda a}^* A_{\lambda a} + 2 A_\chi ^{\alpha *} A_\chi ^{\alpha}
		- 3 M_1 M_1^{\dagger} g_1^{2} - 6 M_L M_L^{\dagger} g_L^{2} 
		+ \frac{3}{8} g_1^{2} {\cal S}_2
	\\
 \nonumber	\\	
16 \pi^2 \deriv{}{t} m_{\bar \chi}^2
	& =	&
		\bar{\lambda}_a \inb[.5cm]
		{
			  4 m_{\bar \chi}^{2} \bar\lambda_a^*
			+ 4 m_{\bar \chi^c}^{2} \bar\lambda_a^*
			+ 4 m_{\Phi a b}^{2} \bar\lambda_b^*
		}
		+ \mu_\chi^{\alpha} \inb
		{
			  2 m_{\bar \chi}^2  \mu_{\chi}^{\alpha *}
			+ 2 m_{\chi}^2 \mu_\chi^{\alpha *}
			+ 2 \inp{m_{S}^2}^{\alpha \beta} \mu_\chi^{\beta *}	
		}
\nonumber	\\
	&	&
		+ \; 4 A_{\bar\lambda a}^* A_{\bar\lambda a}
		+ 2 A_{\chi}^{\alpha *} A_{\chi}^{\alpha}
		- 3 M_1 M_1^{\dagger} g_1^{2} - 6 M_L M_L^{\dagger} g_L^{2} 
		- \frac{3}{8} g_1^{2} {\cal S}_2
	\\
 \nonumber	\\
16 \pi^2 \deriv{}{t} m_{\chi^c}^2
	& =	&
		\lambda_a^{*} \inb[.5cm]
		{
			  4 m_{\chi^c}^{2} \lambda_a
			+ 4 m_{\chi}^{2} \lambda_a
			+ 4 m_{\Phi b a}^{2} \lambda_b
		}
		+ \mu_{\chi^c}^{\alpha *} \inb
		{
			    2 m_{\chi^c}^2 \mu_{\chi^c}^{\alpha}
			  + 2 m_{\bar\chi^c}^2 \mu_{\chi^c}^{\alpha} 
			  + 2 \mu_{\chi^c}^{\beta} \inp{m_{S}^2}^{\beta \alpha}
		}
\nonumber	\\
	&	&
		+ \; 4 A_{\lambda a}^{*} A_{\lambda a} + 2 A_{\chi^c}^{\alpha *} A_{\chi^c}^{\alpha}
		- 3 M_1 M_1^{\dagger} g_1^{2} - 6 M_R M_R^{\dagger} g_R^{2} 
		- \frac{3}{8} g_1^{2} {\cal S}_2
	\\
 \nonumber	\\	
16 \pi^2 \deriv{}{t} m_{\bar \chi^c}^2
	& =	&
		  \bar\lambda_a^{*} \inb[.5cm]
		{
			  4 m_{\bar \chi^c}^{2} \bar\lambda_a
			+ 4 m_{\bar \chi}^{2} \bar\lambda_a
			+ 4 m_{\Phi b a}^{2} \bar\lambda_b	
		}
		+ \mu_{\chi^c}^{\alpha *} \inb
		{
			  2 m_{\bar \chi^c}^2 \mu_{\chi^c}^{\alpha}
			+ 2 m_{\chi^c}^2 \mu_{\chi^c}^{\alpha} 
			+ 2 \mu_{\chi^c}^{\beta} \inp{m_{S}^2}^{\beta \alpha}
		}
\nonumber	\\
	&	&
		+ \; 4 A_{\bar\lambda a}^{*} A_{\bar\lambda a} + 2 A_{\chi^c}^{\alpha *} A_{\chi^c}^{\alpha}
		- 3 M_1 M_1^{\dagger} g_1^{2} - 6 M_R M_R^{\dagger} g_R^{2} 
		+ \frac{3}{8} g_1^{2} {\cal S}_2
	\\
 \nonumber	\\
16\pi^2 \deriv{}{t} \inp{m_S^2}^{\alpha \beta}
	& =	& \inp{m_S^2}^{\alpha \rho}
		\inb
		{
			  2 \mu_{\chi^c}^{\rho*} \mu_{\chi^c}^{\beta} 
			+ 2 \mu_{\chi}^{\rho*} \mu_\chi^{\beta}
			+ \half \conj{\inp{Y^{\rho \mu \nu }}} Y^{\beta \mu \nu}
			+ 8\Tr\inp
				{
					 \mu_\Phi^{\rho \rho} \mu_\Phi^{\beta}
				}			
		}
 \nonumber	\\
	&	&
		+ \inb
		{
			  2 \mu_{\chi^c}^{\alpha *} \mu_{\chi^c}^{\rho}
			+ 2 \mu_\chi^{\alpha *} \mu_{\chi}^{\rho}
			+ \half \conj{\inp{Y^{\alpha \mu \nu }}} Y^{\rho \mu\nu } 
			+ 8 \Tr\inp
				{
					\mu_\Phi^{\alpha \dagger} \mu_\Phi^{\rho }
				}
		}
		\inp{m_S^2}^{\rho \beta}
 \nonumber	\\
	&	&
		+ \; 4 \mu_\chi^{\alpha*} \mu_\chi^{\beta} \inp
		{
			m_{\bar\chi}^2 +m_\chi^2
		}
		+ 4 \mu_{\chi^c}^{\alpha*} \mu_{\chi^c}^{\beta} \inp
		{
			m_{\bar\chi^c}^2 + m_{\chi^c}^2
		}
 \nonumber	\\
	&	&
		+ \; 2 \conj{\inp{Y^{\alpha \rho \mu }}} Y^{\beta \rho \nu} \inp{m_S^2}^{\nu \mu}
		+ 32 \Tr \inp
		{
			\mu_\Phi^\beta m_{\Phi}^2 \mu_{\Phi}^{\alpha \dagger}
		}
		+ 4 A_\chi^{\alpha *} A_\chi^{\beta}
 \nonumber	\\
	&	&
		+ \; 4 A_{\chi^c}^{\alpha *} A_{\chi^c}^{\beta}
		+ \conj{\inp{A_S^{\alpha \mu \nu }}} A_S^{\beta\mu \nu}
		+ 16 \Tr 
		\inp
		{
			A_\Phi^{\alpha \dagger} A_\Phi^{\beta}
		}
	\\
16\pi^2 \deriv{}{t} m_{\Phi a b}^{2}
	& = 	&
		m_{\Phi a c}^2
		\inb
		{
			\Tr \inp
			{
				3 h_c^\dagger h_b + h_c^{\prime \dagger} h_b^\prime
			}
			+ \lambda_c^* \lambda_b + \bar\lambda_c^* \bar\lambda_b
			+ 4 \inp
			{
				\mu_\Phi^{\alpha \dagger} \mu_\Phi^\alpha
			}_{c b}
		}
 \nonumber	\\
	&	&
		+ \inb
		{
			\Tr \inp
			{
				3 h_a^\dagger h_c + h_a^{\prime \dagger} h_c^\prime
			}
			+ \lambda_a^* \lambda_c + \bar\lambda_a^* \bar\lambda_c
			+ 4 \inp
			{
				\mu_\Phi^{\alpha \dagger} \mu_\Phi^\alpha
			}_{a c}
		}
		m_{\Phi c b}^2 
 \nonumber	\\
	&	&
		+ \; \Tr \inb
		{
			6 h_a^\dagger h_b m_{Q^c}^2 + 6 h_a^\dagger m_Q^2 h_b
			+ 2 h_a^{\prime \dagger} h_b^\prime m_{L^c}^2 + 2 h_a^{\prime \dagger} m_L^2 h_b^\prime
			+ 6 A_{Q a}^\dagger A_{Q b} + 2 A_{L a}^\dagger A_{L b}
		}
 \nonumber	\\
	&	&
		+ 
		\inb
		{
			  8 \mu_\Phi^{\alpha \dagger} \conj{\inp{m_\Phi^{2}}} \mu_\Phi^{\alpha}
			+ 8 \mu_\Phi^{\alpha \dagger} \mu_\Phi^\beta \inp{m_S^2}^{ \beta \alpha}
			+ 8 A_\Phi^{\alpha \dagger} A_\Phi^\alpha
			- 6 M_L M_L^\dagger g_L^2 
			- 6 M_R M_R^\dagger g_R^2
		}_{a b}
 \nonumber		\\
	&	&
		+ \; 2 \lambda_a^* \lambda_b
		\inp
		{
			m_{\chi^c}^2 + m_{\chi}^2
		}
		+ 2 \bar\lambda_a^* \bar\lambda_b
		\inp
		{
			m_{\bar\chi^c}^2 + m_{\bar\chi}^2
		}
		+ 2 A_{\lambda a}^* A_{\lambda b}
		+ 2 A_{\bar\lambda a}^* A_{\bar\lambda b}
\end{eqnarray}

\section{Concerning Triplets}
%
\label{Section: Triplets}
%
\subsection{Particle Content \& Quantum Numbers}

\noindent Table \ref{Tbl: Particle Content} shows the various particles of the triplet version of the SUSYLR model and their representations -- except for the $U(1)_{B-L}$ group where the $B - L$ number is given ({The $B - L$ number used in the RGEs follows the GUT normalization scheme; the values in the table do not.  To get the GUT-normalized value, multiply the number in the table by $\sqrt{3/8}$}).

The $Q$ and the $L$ are the standard quarks and leptons of the MSSM while the $Q^c$ and $L^c$ contain the corresponding right-handed conjugate fields.  In order to keep this model general, we allow for an arbitrary number of singlet fields and bidoublet fields.  These values are $n_S$ and $n_\Phi$, respectively.  Thus, for $S^\alpha$, we have $\alpha = 1, 2, \ldots, n_S$; for $\Phi_a$, we have $a = 1, 2, \ldots, n_\Phi$.
(for further comments on $n_\Phi$ see section \ref{Section: Doublets: Particle Content}).

For the following work the particles have been chosen to have the form shown below, where the $Q$ and the $Q^c$ fields serve as templates to construct the other $SU(2)$ doublets (note that the color and generations have been supressed here). The charge is determined by the equation $Q = I_{3L} + I_{3R} + \frac{B - L}{2}$ and the standard $I_3$ ordering is used (row one has the highest $I_3$ value, row two the next highest, etc).

\begin{table}
\begin{center}
\begin{tabular}{lccccccc}
		& $SU(3)^c$	&$\times$& $SU(2)_L$	&$\times$& $SU(2)_R$	&$\times$& $U(1)_{B-L}$		\\
$Q$		& 3		&	 & 2		&	 & 1		&	 & $+\third$		\\
$Q^c$		& 3		&	 & 1		&	 & 2		&	 & $-\third$		\\
$L$		& 1		&	 & 2		&	 & 1		&	 & $-1$			\\
$L^c$		& 1		&	 & 1		&	 & 2		&	 & $+1$			\\
$\Phi_a$	& 1		&	 & 2		&	 & 2		&	 & $0$			\\
$\Delta$	& 1		&	 & 3		&	 & 1		&	 & $+2$			\\
$\Delta^c$	& 1		&	 & 1		&	 & 3		&	 & $-2$			\\
$\bar{\Delta}$	& 1		&	 & 3		&	 & 1		&	 & $-2$			\\
$\bar{\Delta}^c$& 1		&	 & 1		&	 & 3		&	 & $+2$			\\
$S^\alpha$	& 1		&	 & 1		&	 & 1		& 	 & $0$
\end{tabular}
\caption{This table shows the representations for the non-abelian gauge groups and the $B-L$ number for $U(1)$.  The $B - L$ number as presented needs to be normalized; when using the GUT normalization (as this paper does), this means multiplying it by $\sqrt{3/8}$}
\label{Tbl: Particle Content}
\end{center}
\end{table}
%
%
%
%
$$
\begin{array}{ll}
Q = \colTwoVect{u}{d}
&
Q^c = \colTwoVect{d^c}{-u^c}
\\
\Delta = \twoByTwo	{ \frac{\delta^+}{\sqrt2} }	{ \delta^{++} }
			{ \delta^0}			{ - \frac{\delta^+}{\sqrt 2} }
&
\Delta^c = \twoByTwo	{ - \frac{\delta^{c-}}{\sqrt2} }	{ - \delta^{c0} }
			{ - \delta^{c--}}			{ \frac{\delta^{c-}}{\sqrt 2} }
\\
\\
\\
\end{array}
$$
$$
\begin{array}{ll}
\bar{\Delta} = \twoByTwo	{ \frac{\bar{\delta}^{-}}{\sqrt2} }		{ \bar{\delta}^{0} }
				{ \bar{\delta}^{--}}				{ - \frac{\bar{\delta}^{-}}{\sqrt 2} }
&
\bar{\Delta}^c = \twoByTwo
	{ - \frac{\bar{\delta}^{c+}}{\sqrt2} }	{ - \bar{\delta}^{c++} }
	{ - \bar{\delta}^{c0}}			{ \frac{\bar{\delta}^{c+}}{\sqrt 2} }
\\
\multicolumn{2}{c}{
\Phi_a = \twoByTwo	{\nop{\phi_d^0}_a}		{\nop{\phi_u^+}_{\!\! a}}
			{\nop{\phi_d^-}_{\!\! a}}	{\nop{\phi_u^0}_a}
}
\end{array}
$$

%
\noindent These fields transform under $SU(2)$ as
%
%
%
$$
\begin{array}{ccccccc}
Q \rightarrow U_L Q						& & 
Q^c \rightarrow U_R Q^c						\\
L \rightarrow U_L L						& & 
L^c \rightarrow U_R L^c						\\
\Delta \rightarrow U_L \Delta U_L^\dagger			& &
\Delta^c \rightarrow U_R \Delta^c U_R^\dagger			\\
\bar{\Delta} \rightarrow U_L \bar{\Delta} U_L^\dagger		& &
\bar{\Delta}^c \rightarrow U_R \bar{\Delta}^c U_R^\dagger	\\
\Phi_a  \rightarrow U_L \Phi_a U_R^\dagger			& &
S^{\alpha} \rightarrow S^{\alpha}
\end{array}
$$
and under Parity as
%
%
%
$$
\begin{array}{ccccccc}
Q \rightarrow -i \tau_2 Q^{c \, *}				& & 
Q^c \rightarrow i \tau_2 \conj{Q}				\\ 
L \rightarrow -i \tau_2 L^{c \, *}				& & 
L^c \rightarrow i \tau_2 \conj{L}				\\
\Delta \rightarrow \tau_2 \Delta^{c \, *} \tau_2		& &
\Delta^c \rightarrow \tau_2 \conj{\Delta} \tau_2		\\
\bar{\Delta} \rightarrow \tau_2 \bar{\Delta}^{c \, *} \tau_2	& &
\bar{\Delta}^c \rightarrow \tau_2 \conj{\bar{\Delta}} \tau_2	\\
\Phi_a  \rightarrow \Phi_a^\dagger				& &
S^{\alpha} \rightarrow S^{\alpha \, *}
\end{array}
$$
%



\subsection{Superpotential and Soft Breaking Lagrangian}

With the transformations and representations given above, the most general superpotential and soft breaking terms are
%
%
\begin{eqnarray}
\nonumber
W	& =	&   i h_a Q^T \tau_2 \Phi_a Q^c
		  + i h_a^\prime L^T \tau_2 \Phi_a L^c
		  + i f L^T \tau_2 \Delta L
		  + i f_c L^{cT} \tau_2 \Delta^c L^c
	\\ \nonumber
	& 	& + \; M_\Delta \Tr\inp{\Delta \bar{\Delta}}
		  + M_{\Delta^c} \Tr\inp{\Delta^c \bar{\Delta}^c}
		  + M_{\Phi a b} \Tr\inp{\Phi_a^T \tau_2 \Phi_b \tau_2}
	\\ \nonumber
	&	& + \; \mu^\alpha_\Delta S^\alpha \Tr\inp{\Delta \bar{\Delta}}
		  + \mu^\alpha_{\Delta^c}  S^\alpha \Tr\inp{\Delta^c \bar{\Delta}^c}
		  + \mu^\alpha_{\Phi a b } S^\alpha \Tr\inp{\Phi_a^T \tau_2 \Phi_b \tau_2}
	\\
	&	& + \; \sixth Y^{\alpha \beta \gamma} S^\alpha S^\beta S^\gamma
		  + \half M_S^{\alpha \beta} S^\alpha S^\beta
		  + L^\alpha S^\alpha
\end{eqnarray}
%
%
%
\begin{eqnarray}
\nonumber
- {\cal L}_{SB}
	& =	& \half \inp{	  M_3 \superp{g} \superp{g} 
				+ M_L \superp{W_L} \superp{W_L}
				+ M_R \superp{W_R} \superp{W_R}
				+ M_1 \superp{B} \superp{B}
				+ \mbox{h.c.}
			    }
	\\ \nonumber
	&	& + \left[ \parbox[h][0.75cm]{0cm}{}
		    i A_{Q a} \superp{Q}^T \tau_2 \Phi_a \superp{Q^c}
		  + i A_{L a} \superp{L}^T \tau_2 \Phi_a \superp{L^c}
		  + i A_{f} \superp{L}^T \tau_2 \Delta \superp{L}
		  \right.
	\\ \nonumber
	& 	& \indent
		  + i A_{f^c} \superp{L^c}^T \tau_2 \Delta^c \superp{L^c}
		  + A^\alpha_{\Delta} S^\alpha \Tr \inp{ \Delta \bar{\Delta} }
		  + A^\alpha_{\Delta^c} S^\alpha \Tr \inp{ \Delta^c \bar{\Delta}^c }
	\\ \nonumber
	&	& \left. \indent
		  + A^\alpha_{\Phi a b} S^\alpha \Tr \inp{ \Phi_a^T \tau_2 \Phi_b \tau_2 }
		  + \sixth A_S^{\alpha \beta \gamma} S^\alpha S^\beta S^\gamma 
		  + \mbox{h.c.}
		  \parbox[h][0.75cm]{0cm}{}
		  \right]
	\\ \nonumber
	&	& + \left[ \parbox[h][0.75cm]{0cm}{}
		    B_\Delta \Tr \inp{ \Delta \bar{\Delta} }
		  + B_{\Delta^c} \Tr \inp{ \Delta^c \bar{\Delta}^c }
		  + B_{\Phi a b} \Tr \inp{ \Phi_a^T \tau_2 \Phi_b \tau_2 }
		  + \half B_S^{\alpha \beta} S^\alpha S^\beta
		  + \mbox{h.c.}
		  \parbox[h][0.75cm]{0cm}{}
		  \right]
	\\ \nonumber
	&	& + \left[ \parbox[h][0.75cm]{0cm}{}
		    m_Q^2 \superp{Q}^T \conj{\superp{Q}}
		  + m_{Q^c}^2 \superp{Q^c}^\dagger \superp{Q^c}
		  + m_{L}^2 \superp{L}^T \conj{\superp{L}}
		  + m_{L^c}^2 \superp{L^c}^\dagger \superp{L^c}
		  \right.
	\\ \nonumber
	&	& \indent
		  + m_{\Delta}^2 \Tr \inp{ \Delta^\dagger \Delta }
		  + m_{\bar{\Delta}}^2 \Tr \inp{ \bar{\Delta}^\dagger \bar{\Delta} }
		  + m_{\Delta^c}^2 \Tr \inp{ \Delta^{c \, \dagger} \Delta^c }
		  + m_{\bar{\Delta}^c}^2 \Tr \inp{ \bar{\Delta}^{c \, \dagger} \bar{\Delta}^c }
	\\
	&	&
		  \left. \indent
		  + m_{\Phi ab}^2 \Tr \inp{ \Phi_a^\dagger \Phi_b }
		  + \inp{m_S^2}^{\alpha \beta} \conj{\inp{S^\alpha}} S^\beta
		  \parbox[h][0.75cm]{0cm}{}
		  \right]
\end{eqnarray}
where the generational and color indices have been supressed and the transposes and $\tau_2$'s belong to $SU(2)$.  Thus the first term in $W$ is actually 
$$
i \inp{h_a}^{j}_{\;k} \rowTwoVect{u_{jA}}{d_{jA}} \tau_2 \Phi_a \colTwoVect{ d^{c\,k}_A }{ - u^{c\,k}_A}
$$
with the lowercase latin indicies specifying the generation and the uppercase latin indices specifying color.

By demanding parity invariance from this theory, we also find the following relations among the parameters \cite{Journal:Phys.Rev.Lett.76:3490.1996, Journal:Phys.Rev.Lett.76:3486.1996, Journal:Phys.Rev.Lett.79:4744.1997}: soft breaking mass terms for an $SU(2)_L$ doublet are equal to those of the corresponding $SU(2)_R$ doublet and
$$
\begin{array}{*{7}{c}}
h_a = h_a^\dagger					&	&
h_a^\prime = h_a^{\prime \dagger}			&	&
f = f_c^*						&	&
\mu_\Delta^\alpha = \mu_{\Delta^c}^{\alpha *}		\\
\mu_{\Phi ab} = \mu_{\Phi ab}^*				&	&
M_\Delta = M_{\Delta^c}^*				&	&
M_{\Phi a b} = M_{\Phi a b}^*				&	&
M_S^{\alpha \beta} = M_S^{\alpha \beta *}
							\\
L^\alpha = L^{\alpha *}					&	&
M_1 = M_1^*						&	&
M_L = M_R^*						&	&
M_3 = M_3^*						\\
g_L = g_R						&	&
B_\Delta = B_{\Delta^c}^*				&	&
B_{\Phi a b} = B_{\Phi a b}^*				&	&
B_S^{\alpha \beta} = B_S^{\alpha \beta *}
\end{array}
$$
Where $g_L$ and $g_R$ are the $SU(2)_L$ and $SU(2)_R$ coupling constants, respectively.

\subsection{RGEs}
%
The renormalization group equations to one-loop order for all the parameters of the above theory are presented below and are categorized by the type of coupling
%
%
\subsubsection{Gauge Couplings}
%
\vspace{-.5cm}
\begin{equation}
\begin{array}{*{3}{c}}
\displaystyle
16\pi^2 \deriv{}{t} g_1		=	24 g_1^{3}			&	& \displaystyle
16\pi^2 \deriv{}{t} g_L		= 	\inp{4 + n_\Phi} g_L^{3}	\\ \displaystyle
16\pi^2 \deriv{}{t} g_R		= 	\inp{4 + n_\Phi} g_R^{3}	&	& \displaystyle
16\pi^2 \deriv{}{t} g_3		= 	-3 g_3^{3}
\end{array}
\end{equation}

%
%
\subsubsection{Yukawa Couplings}
%
\vspace{-.5cm}
%
%
\begin{eqnarray}
\nonumber
16\pi^2 \deriv{}{t} h_a
	& =	&  h_a \left[
			  2 h_b^\dagger h_b 
		   	- \sixth g_1^2
			- 3 g_R^2 
			- 3 g_L^3
			- \frac{16}{3} g_3^2
			\right]
	\\
	&	&
		  + \; h_b \inb{	  \Tr\inp{3 h_b^\dagger h_a + h_b^{\prime\dagger} h_a^\prime}
					+ 2 h_b^\dagger h_a
					+ 4 \inp{ \mu_\Phi^{\alpha \, \dagger} \mu_\Phi^\alpha }_{ba}
				}
\end{eqnarray}


\begin{eqnarray}
\nonumber
16\pi^2 \deriv{}{t} h_a^\prime
	& =	&   h_a^\prime \inb{	6 f_c^\dagger f_c 
					+ 2 h_b^{\prime \dagger} h_b^{\prime} 
					- \frac{3}{2} g_1^2 
					- 3 g_R^2
					- 3 g_L^2
				  }
		  + 6 f f^\dagger h_a^\prime
	\\
	&	&
		  + \; h_b^\prime \inb{	  2 h_b^{\prime \dagger} h_a^{\prime}
					+ \Tr\inp{3 h_b^\dagger h_a + h_b^{\prime\dagger} h_a^\prime}
					+ 4 \inp{ \mu_\Phi^{\alpha \, \dagger} \mu_\Phi^\alpha }_{ba}
		    			}
\end{eqnarray}

%
\begin{eqnarray}
\nonumber
16 \pi^2 \deriv{}{t} f
	& =	&  f \inb{	6 f^\dagger f 
				+ 2 h_a^{\prime \, *} h_a^{\prime \, T} 
		  		+ 2 \Tr\inp{f^\dagger f} 
				+ \mu_\Delta^{\alpha \, *} \mu_\Delta^\alpha
				- \frac{9}{2} g_1^2 - 7 g_L^2
			}
	\\
	&	&
		  + \; \inb{ 6 f f^{\dagger} + 2 h_a^\prime h_a^{\prime \dagger} } f
\end{eqnarray}
%
%
\begin{eqnarray}
\nonumber
16 \pi^2 \deriv{}{t} f_c
	& =	&  f_c \inb{ 	6  f_c^\dagger f_c 
				+ 2 h_a^{\prime \dagger} h_a^{\prime}
				+ 2 \Tr\inp{f_c^\dagger f_c}
		  		+ \mu_{\Delta^c}^{\alpha \, *} \mu_{\Delta^c}^\alpha
		  		- \frac{9}{2} g_1^2 
		  		- 7 g_R^2
			   }
	\\
	&	&
		  + \inb{	6 f_c f_c^\dagger 
		  		+ 2 h_a^{\prime \, T} h_a^{\prime \, *} 
		  	} f_c
\end{eqnarray}
%
%
\begin{eqnarray}
\nonumber
16 \pi^2 \deriv{}{t}\mu_{\Delta}^\alpha
	& =	& \mu_\Delta^\alpha \inb{	2 \Tr\inp{ f^{\dagger} f }
						+ 2 \mu_{\Delta}^{\beta \, *} \mu_{\Delta}^\beta
						- 6 g_1^2
						- 8 g_L^2
					  }
	\\
	&	&
		  + \; \mu_{\Delta}^\beta
		  	\inb{ \YSmnYSmn{\beta}{\alpha} }
\end{eqnarray}
%
%
\begin{eqnarray}
\nonumber
16 \pi^2 \deriv{}{t} \mu_{\Delta^c}^\alpha
	& =	& \mu_{\Delta^c}^\alpha
			\inb{	2 \Tr\inp{f_c^\dagger f_c}
				+ 2 \mu_{\Delta^c}^{\beta \, *} \mu_{\Delta^c}^\beta
				- 6 g_1^2
				- 8 g_R^2
			    }
	\\
	&	& + \; \mu_{\Delta^c}^\beta
		  	\inb{  \YSmnYSmn{\beta}{\alpha}  }
\end{eqnarray}
%
%
\begin{eqnarray}
\nonumber
16 \pi^2 \deriv{}{t} \mu_{\Phi ab}^\alpha
	& =	& \mu_{\Phi ac}^\alpha
		  	\left[	\Tr\inp{ 3 h_c^{\dagger} h_b +  h_c^{\prime \, \dagger} h_b^\prime }
		  		+ 4 \inp{ \mu_\Phi^{\beta \, \dagger} \mu_\Phi^{\beta} }_{cb}
		  	\right]
	\\ \nonumber
	&	&
		  + \left[	\Tr\inp{ 3 h_a h_c^{\dagger} +  h_a^\prime h_c^{\prime \, \dagger}  } 
				+ 4 \inp{ \mu_\Phi^\beta \mu_\Phi^{\beta \, \dagger} }_{ac}
			\right] \mu_{\Phi cb}^\alpha
	\\ \nonumber
	&	& 
		  + \;  \mu_{\Phi ab}^\beta \inb{	  \YSmnYSmn{\beta}{\alpha}
		  				}
	\\
	&	& 
		  + \; \mu_{\Phi ab}^\alpha \inb{ - 3g_L^2 - 3 g_R^2 }
\end{eqnarray}
%
%
\begin{eqnarray}
\nonumber
16 \pi^2 \deriv{}{t} Y^{\alpha \beta \gamma}
	& =	& Y^{\alpha \beta \rho} \inb{ \YSmnYSmn{\rho}{\gamma} }
	\\ \nonumber
	& 	& + \; Y^{\gamma \beta \rho} \inb{ \YSmnYSmn{\rho}{\alpha} }
	\\
	& 	& + \; Y^{\alpha \gamma \rho} \inb{ \YSmnYSmn{\rho}{\beta} }
\end{eqnarray}
%
%
%
%
\subsubsection{Mass Terms}
%
\vspace{-.5cm}
%
%
\begin{eqnarray}
16 \pi^2 \deriv{}{t} M_\Delta
	& =	& M_\Delta
			\inb{	2 \Tr\inp{f^\dagger f}
				+ 2 \mu_\Delta^{\alpha \, *} \mu_\Delta^{\alpha}
				- 6 g_1^2 
				- 8 g_L^2
			    }
\end{eqnarray}
%
%
\begin{eqnarray}
16 \pi^2 \deriv{}{t} M_{\Delta^c}
	& =	& M_{\Delta^c}
			\inb{	2 \Tr\inp{f_c^\dagger f_c}
				+ 2 \mu_{\Delta^c}^{\alpha \, *} \mu_{\Delta^c}^\alpha
				- 6 g_1^2 
				- 8 g_R^2
			    }
\end{eqnarray}
%
%
\begin{eqnarray}
\nonumber
16 \pi^2 \deriv{}{t} M_{\Phi ab}
	& =	&    M_{\Phi ac}
			\inb{	 \Tr\inp{ 3 h_c^{\dagger} h_b + h_c^{\prime \, \dagger} h_b^\prime } 
		 		 + 4 \inp{\mu_\Phi^{\beta \, \dagger} \mu_\Phi^\beta }_{cb}
			    }
		  + M_{\Phi ab} \inp{-6 g_L^2 - 6 g_R^2}
	\\ 
	&	& + \inb{	\Tr\inp{ 3 h_a h_c^{\dagger} +  h_a^\prime h_c^{\prime \, \dagger}  } 
				+ 4 \inp{ \mu_\Phi^\beta \mu_{\Phi}^{\beta \, \dagger}}_{ac}
		     	 } M_{\Phi cb}
\end{eqnarray}
%
%
\begin{eqnarray}
\nonumber
16 \pi^2 \deriv{}{t} M_S^{\alpha \beta}
	& =	& M_S^{\alpha \rho} \inb{ \YSmnYSmn{\rho}{\beta} }
	\\
	&	& + \; M_S^{\beta \rho} \inb{ \YSmnYSmn{\rho}{\alpha} }
\end{eqnarray}
%
%
%
\subsubsection{Linear Term}
%
\vspace{-.5cm}
%
%
\begin{eqnarray}
16 \pi^2 \deriv{}{t} L^{\alpha}
	& =	& L^{\rho} \inb{ \YSmnYSmn{\rho}{\alpha} }
\end{eqnarray}
%
%
\subsubsection{Gaugino Masses}
%
\vspace{-.5cm}
\begin{equation}
\begin{array}{*{3}{c}}
\displaystyle
	16\pi^2 \deriv{}{t} M_1	=	48 M_1 g_1^{2}				&	& \displaystyle
	16\pi^2 \deriv{}{t} g_L	=	\inp{8 + 2 n_\Phi} M_L g_L^{2}		\\ \nonumber \displaystyle
	16\pi^2 \deriv{}{t} g_R	=	\inp{8 + 2 n_\Phi} M_R g_R^{2}		&	& \displaystyle
	16\pi^2 \deriv{}{t} g_3	=	-6 M_3 g_3^{2}
\end{array}
\end{equation}
%
%
\subsubsection{Soft Breaking Trilinear $A$'s}
%
\vspace{-.5cm}
\begin{eqnarray}
\nonumber
16 \pi^2 \deriv{}{t} A_{Q a}
	& =	& A_{Q a} \inb{	  2 h_b^\dagger h_b
				- \frac{1}{6} g_1^2
				- 3 g_L^2 
				- 3 g_R^2
				- \frac{16}{3} g_3^2
			      }
		  + 2 h_b h_b^\dagger A_{Q a}
	\\ \nonumber
	&	& + \; h_a \inb{	  4 h_b^\dagger A_{Q b}
		  			+ \frac{1}{3} g_1^2 M_1
		  			+ 6 g_L^2 M_L
		  			+ 6 g_R^2 M_R
		  			+ \frac{32}{3} g_3^2 M_3
		  		}
		  + 4 A_{Q b} h_b^\dagger h_a
	\\ \nonumber
	&	&
		  + \inb{	  \Tr\inp{3 h_a h_b^\dagger + h_a^\prime h_b^{\prime\dagger}}
				+ 4 \inp{ \mu_\Phi^\alpha \mu_\Phi^{\alpha \, \dagger} }_{ab}
			} A_{Q b}
	\\
	&	&
		  + \inb{	  \Tr\inp{6 A_{Q a} h_b^\dagger + 2 A_{L a} h_b^{\prime\dagger}}
				+ 8 \inp{ A_\Phi^\alpha \mu_\Phi^{\alpha \, \dagger} }_{ab}
			} h_b
\end{eqnarray}
%
%
\begin{eqnarray}
\nonumber
16 \pi^2 \deriv{}{t} A_{L a}
	& =	& A_{L a} \inb{	6 f_c^\dagger f_c 
		  		+ 2 h_b^{\prime \dagger} h_b^{\prime}
				- \frac{3}{2} g_1^2 
				- 3 g_L^2
				- 3 g_R^2
				}
	\\ \nonumber
	&	&
		  + \; h_a^\prime \inb{	12 f_c^\dagger A_{f_c}
		 			+ 4 h_b^{\prime \dagger} A_{L b}
					+ 3 g_1^2 M_1
					+ 6 g_L^2 M_L
					+ 6 g_R^2 M_R
				    }
	\\ \nonumber
	&	& + \inb{	6 f f^{\dagger}
			  	+ 2 h_b^\prime h_b^{\prime \dagger}
			} A_{L a}
		  + \inb{	12 A_f f^{\dagger}
			  	+ 4 A_{L b} h_b^{\prime \dagger}
			    } h_a^\prime
	\\ \nonumber
	&	&
		  + \; A_{L b}
			\inb{
				\Tr\inp{3 h_b^\dagger h_a + h_b^{\prime\dagger} h_a^\prime}
			 	+ 4 \inp{\mu_\Phi^{\alpha \, \dagger} \mu_\Phi^\alpha }_{ba}
			    }
	\\
	&	&
		  + \; h_b^\prime 
			\inb{
				\Tr\inp{6 h_b^\dagger A_{Q a} + 2 h_b^{\prime\dagger} A_{L a}}
			 	+ 8 \inp{\mu_\Phi^{\alpha \, \dagger} A_\Phi^\alpha }_{ba}
			    }
\end{eqnarray}
%
%
\begin{eqnarray}
\nonumber
16 \pi^2 \deriv{}{t} A_f 
 	& =	&  A_f \inb{ 	  6 f^\dagger f 
 				+ 2 h_a^{\prime \, *} h_a^{\prime \, T}
		  		+ 2 \Tr\inp{f^\dagger f}
		  		+ \mu_\Delta^{\alpha \, *} \mu_\Delta^\alpha
				- \frac{9}{2} g_1^2
				- 7 g_L^2
			    }
	\\ \nonumber
	&	&
		  +\; f	\left[ 	  12 \conj{f} A_f 
		  		+ 4 h_a^{\prime \, *} A_{L a}^T 
				+ 4 \Tr\inp{f^\dagger A_f}
		  		+ 2 \mu_\Delta^{\alpha \, *} A_\Delta^\alpha
		  		+ 9 g_1^2 M_1
				+ 14 g_L^2 M_L
			\right]
	\\
	&	& + \inb{	  6 f f^{\dagger}
				+ 2 h_a^\prime h_a^{\prime \dagger}
			    } A_f
		  +  \inb{	  12 A_f f^{\dagger}
				+ 4 A_{L a} h_a^{\prime \dagger}
			    } f
\end{eqnarray}
%
%
\begin{eqnarray}
\nonumber
16 \pi^2 \deriv{}{t} A_{f^c}
	& =	&   A_{f^c} \inb{	  6 f_c^\dagger f_c
		  			+ 2 h_a^{\prime \dagger} h_a^{\prime}
					+ 2 \Tr\inp{f_c^\dagger f_c}
		  			+ \mu_{\Delta^c}^{\alpha \, *} \mu_{\Delta^c}^\alpha
		  		 	- \frac{9}{2} g_1^2
					- 7 g_R^2
				}
	\\ \nonumber
	&	& + \; f_c 	\left[	  12 f_c^\dagger A_{f^c}
		  			+ 4 h_a^{\prime \dagger} A_{L a}
		  			+ 4 \Tr\inp{f_c^\dagger A_{f^c}}
		  			+ \; 2 \mu_{\Delta^c}^{\alpha \, *} A_{\Delta^c}^\alpha
		  			+ 9 g_1^2 M_1
					+ 14 g_R^2 M_R
			    	\right]
	\\
	&	& + \inb{	 6 f_c f_c^\dagger  
		  		+ 2 h_a^{\prime \, T} h_a^{\prime \, *}
		  	} A_{f^c}
		  + \inb{	  12 A_{f^c} f_c^\dagger
		  		+ 4 A_{L a}^T h_a^{\prime \, *}
		  	} f_c
\end{eqnarray}
%
%
\begin{eqnarray}
\nonumber
16 \pi^2 \deriv{}{t} A_\Delta^\alpha
	& =	&  A_\Delta^\alpha \inb{	2 \Tr\inp{ f^{\dagger} f }
		  				+ 2 \mu_\Delta^{\beta \, *} \mu_\Delta^\beta
		  				- 6 g_1^2
						- 8 g_L^2
					}
	\\ \nonumber
	&	& + \; \mu_\Delta^\alpha \inb{	  4 \Tr\inp{ f^{\dagger} A_f }
		  				+ 4 \mu_\Delta^{\beta \, *} A_\Delta^\beta
		  			  	+ 12 g_1^2 M_1
						+ 16 g_L^2 M_L
					}
	\\ \nonumber
	&	& + \; A_\Delta^\beta \inb{ \YSmnYSmn{\beta}{\alpha} }
	\\
	&	& + \; \mu_\Delta^\beta \inb{ \YSmnhSmn{\beta}{\alpha} }
\end{eqnarray}
%
%
\begin{eqnarray}
\nonumber
16 \pi^2 \deriv{}{t} A_{\Delta^c}^\alpha
	& =	&  A_{\Delta^c}^\alpha \inb{	2 \Tr\inp{ f_c^{\dagger} f_c }
		  				+ 2 \mu_{\Delta^c}^{\beta \, *} \mu_{\Delta^c}^\beta
		  				- 6 g_1^2
						- 8 g_R^2
					  }
	\\ \nonumber
	&	&
		  + \mu_{\Delta^c}^\alpha \inb{	  4 \Tr\inp{ f_c^{\dagger} A_{f^c} }
		  				+ 4 \mu_{\Delta^c}^{\beta \, *} A_{\Delta^c}^\beta
		  			  	+ 12 g_1^2 M_1
						+ 16 g_R^2 M_R
					}
	\\ \nonumber
	&	&
		  + A_{\Delta^c}^\beta \inb{ \YSmnYSmn{\beta}{\alpha} }
	\\
	&	&
		  + \mu_{\Delta^c}^\beta \inb{ \YSmnhSmn{\beta}{\alpha} }
\end{eqnarray}
%
%
\begin{eqnarray}
\nonumber
16 \pi^2 \deriv{}{t} A_{\Phi ab}^\alpha
	& =	&   A_{\Phi ac}^\alpha
			\inb{	  \Tr\inp{ 3 h_c^{\dagger} h_b +  h_c^{\prime \, \dagger} h_b^\prime }
		  		+ 4 \inp{ \mu_\Phi^{\beta \, \dagger} \mu_\Phi^\beta }_{cb}
			    }
	\\ \nonumber
	&	&
		  + \; \mu_{\Phi ac}^\alpha
			\inb{	  \Tr\inp{ 6 h_c^{\dagger} A_{Q b} +  2 h_c^{\prime \, \dagger} A_{L b} }
		  		+ 8 \inp{ \mu_\Phi^{\beta \, \dagger} A_\Phi^\beta }_{cb}
		  	    }
	\\ \nonumber
	&	&
		  + \inb{	  \Tr\inp{ 3 h_a h_c^{\dagger} +  h_a^\prime h_c^{\prime \, \dagger} }
				+ 4 \inp{ \mu_\Phi^\beta \mu_\Phi^{\beta \, \dagger}  }_{ac}
			} A_{\Phi cb}^\alpha 
	\\ \nonumber
	&	&
		  + \inb{	  \Tr\inp{ 6 A_{Q a} h_c^{\dagger} +  2 A_{L a} h_c^{\prime \, \dagger} }
				+ 8 \inp{ A_\Phi^\beta \mu_\Phi^{\beta \, \dagger} }_{ac}
			} \mu_{\Phi cb}^\alpha
	\\ \nonumber
	&	& + \; A_{\Phi ab}^\alpha \inb{	- 3 g_L^2
						- 3 g_R^2
						}
		  + \mu_{\Phi ab}^\alpha \inb{	  6 g_L^2 M_L
						+ 6 g_R^2 M_R
					     }
	\\ \nonumber
	&	& + \; A_{\Phi ab}^\beta \inb{ \YSmnYSmn{\beta}{\alpha} }
	\\
	&	&
		  + \; \mu_{\Phi ab}^\beta \inb{ \YSmnhSmn{\beta}{\alpha} }
\end{eqnarray}
%
%
\begin{eqnarray}
\nonumber
16 \pi^2 \deriv{}{t} A_S^{\alpha \beta \gamma}
	& =	& A_S^{\alpha \beta \rho} \inb{ \YSmnYSmn{\rho}{\gamma} }
	\\ \nonumber
	&	&
		+ \; Y^{\alpha \beta \rho} \inb{ \parbox[h][.75cm]{0cm}{} \YSmnhSmn{\rho}{\gamma} }
	\\ \nonumber
	&	&
		+ \; A_S^{\gamma \beta \rho} \inb{ \YSmnYSmn{\rho}{\alpha} }
	\\ \nonumber
	&	&
		+ \; Y^{\gamma \beta \rho} \inb{ \parbox[h][.75cm]{0cm}{} \YSmnhSmn{\rho}{\alpha} }
	\\ \nonumber
	&	&
		+ \; A_S^{\alpha \gamma \rho} \inb{ \YSmnYSmn{\rho}{\beta} }
	\\
	&	&
		+ \; Y^{\alpha \gamma \rho} \inb{ \parbox[h][.75cm]{0cm}{} \YSmnhSmn{\rho}{\beta} }
\end{eqnarray}
\pagebreak
%
%
\subsubsection{Soft Breaking Bilinear $B$'s}
%
\vspace{-.5cm}
%
\hide
{
\begin{eqnarray*}
16 \pi^2 \deriv{}{t} b^{ \field{i} \field{j} }
	& =	&   \half b^{\field{i} p} Y_{pmn} Y^{mn \field{j}}
		  + \half Y^{\field{i} \field{j} p} Y_{pmn} b^{mn}
		  + \mu^{\field{i} p} Y_{pmn} h^{mn \field{j}}
	\\
	&	& - \; 2 b^{ \field{i} \field{j} } \sum_\aleph g_\aleph^2 C_\aleph\inp{\field{i}}
	\\
	&	& + \; 4 \mu^{ \field{i} \field{j} } \sum_\aleph M_\aleph g_\aleph^2 C_\aleph\inp{\field{i}}
	\\
	& 	& + \; \half b^{\field{j} p} Y_{pmn} Y^{mn \field{i}}
		  + \half Y^{\field{j} \field{i} p} Y_{pmn} b^{mn}
		  + \mu^{\field{j} p} Y_{pmn} h^{mn \field{i}}
	\\
	&	& - \; 2 b^{ \field{j} \field{i} } \sum_\aleph g_\aleph^2 C_\aleph\inp{\field{j}}
	\\
	&	& + \; 4 \mu^{ \field{j} \field{i} } \sum_\aleph M_\aleph g_\aleph^2 C_\aleph\inp{\field{j}}
\end{eqnarray*}
}
%
%
\begin{eqnarray}
\nonumber
16 \pi^2 \deriv{}{t} B_\Delta
	& =	&   B_\Delta \inb{ 	  2 \Tr \inp{f^\dagger f} 
					+ 2 \mu_\Delta^{\alpha \, *} \mu_\Delta^\alpha
					- 6 g_1^2
					- 8 g_L^2
				}
	\\ \nonumber
	&	&
		  + \; M_\Delta \inb{	  4 \Tr\inp{f^\dagger A_f}
		  			+ 4 \mu_\Delta^{\alpha \, *} A_\Delta^\alpha
		  			+ 12 g_1^2 M_1
					+ 16 g_L^2 M_L
				 }
	\\
	&	&
		  + \; \mu_\Delta^\alpha \inb{ \YSmnbmn{\alpha} }
\end{eqnarray}
%
%
\begin{eqnarray}
\nonumber
16 \pi^2 \deriv{}{t} B_{\Delta^c}
	& =	&   B_{\Delta^c}
				\inb{ 	  2 \Tr \inp{f_c^\dagger f_c} 
					+ 2 \mu_{\Delta^c}^{\alpha \, *} \mu_{\Delta^c}^\alpha
					- 6 g_1^2
					- 8 g_R^2
				    }
	\\ \nonumber
	&	&
		  + \; M_{\Delta^c}
		  		\inb{	  4 \Tr\inp{f_c^\dagger A_{f^c}}
		  			+ 4 \mu_{\Delta^c}^{\alpha \, *} A_{\Delta^c}^\alpha
		  			+ 12 g_1^2 M_1
					+ 16 g_R^2 M_R
				    }
	\\
	&	&
		  + \; \mu_{\Delta^c}^\alpha \inb{ \YSmnbmn{\alpha} }
\end{eqnarray}
%
%
\begin{eqnarray}
\nonumber
16 \pi^2 \deriv{}{t} B_{\Phi ab}
	& =	&   B_{\Phi ac} \inb{ \YPhimnYPhimn[\alpha]{c}{b} }
	\\ \nonumber
	& 	& + \; M_{\Phi ac} \inb{ \YPhimnhPhimn[\alpha]{c}{b} }
	\\ \nonumber
	&	& + \inb{ \YPhimnYPhimnTranspose[\alpha]{c}{a} } B_{\Phi cb}
	\\ \nonumber
	&	& + \inb{ \YPhimnhPhimnTranspose[\alpha]{c}{a} } M_{\Phi cb}
	\\ \nonumber
	& 	& + \; \mu_{\Phi ab}^\rho \inb{ \YSmnbmn{\rho} }
	\\
	& 	& + \; B_{\Phi ab} \inb{	- 3 g_L^2
						- 3 g_R^2
					}
		  + \; M_{\Phi ab} \inb{	  6 g_L^2 M_L
						+ 6 g_R^2 M_R
					}
\end{eqnarray}
%
%
\begin{eqnarray}
\nonumber
16 \pi^2 \deriv{}{t} B_S^{\alpha \beta}
	& =	&   B_S^{\alpha \rho} \inb{ \YSmnYSmn{\rho}{\beta} }
	\\ \nonumber
	& 	& + \; M_S^{\alpha \rho} \inb{ \YSmnhSmn{\rho}{\beta} \parbox[h][.75cm]{0cm}{} }
	\\ \nonumber
	&	& + \; B_S^{\beta \rho} \inb{ \YSmnYSmn{\rho}{\alpha} }
	\\ \nonumber
	&	& + \; M_S^{\beta \rho} \inb{  \YSmnhSmn{\rho}{\alpha} \parbox[h][.75cm]{0cm}{} }
	\\
	&	& + \; Y^{\alpha \beta \rho} \inb{ \YSmnbmn{\rho} \parbox[h][.75cm]{0cm}{} }
\end{eqnarray}
%
%
\hide
{
\begin{eqnarray*}
16 \pi^2 \deriv{}{t} \inp{ m^2 }^{ \field{j} }_{\;\; \field{i} }
	& =	&  \half Y_{\field{i} pq} Y^{pqn} \inp{m^2}^{\field{j}}_{\;\; n}
		  + \half Y^{\field{j} pq} Y_{pqn} \inp{m^2}^{n}_{\;\; \field{i} }
	\\
	&	& + \; 2 Y_{\field{i} pq} Y^{\field{j} pr} \inp{m^2}^{q}_{\;\; r}
		  + h_{\field{i} pq} h^{\field{j} pq}
	\\
	& 	& - \; 8 \delta^{\field{i}}_{\;\; \field{j}} 
					\sum_\aleph M_\aleph M_\aleph^\dagger g_\aleph^2 C_\aleph\inp{\field{i}}
	\\
	&	& + \; 2 g_1^2 \inp{ t_{B - L}^{A} }^{\field{j}}_{\;\; \field{i}} \Tr\inp{t_{B-L}^A m^2}
\end{eqnarray*}
}
%
\subsubsection{Soft Breaking Masses}
%
Since each of the RGEs for the soft breaking masses have the following term in common, it is convenient to define
\begin{eqnarray*}
\massSum[S]	& \equiv	& \massSum
\end{eqnarray*}
%
%
%
\begin{eqnarray}
\nonumber
16 \pi^2 \deriv{}{t} m_Q^2
	& =	&   2 m_Q^2 h_a h_a^\dagger 
		  + h_a \inb[.75cm]{	  2 h_a^\dagger m_Q^2 
			  		+ 4 h_b^\dagger m_{\Phi ab}^2
			  		+ 4 m_{Q^c}^2 h_a^\dagger
		  		     }
	\\
	&	& + \; 4 A_{Q a} A_{Q a}^\dagger
		  - \frac{1}{3} M_1 M_1^\dagger g_1^2
		  - 6 M_L M_L^\dagger g_L^2
		  - \frac{32}{3} M_3 M_3^\dagger g_3^2
		  + \eighth g_1^2 \massSum[S]
\end{eqnarray}
%
%
%
%
%
\begin{eqnarray}
\nonumber
16 \pi^2 \deriv{}{t} m_{Q^c}^2
	& =	&   2  m_{Q^c}^2 h_a^\dagger h_a 
		  + h_a^\dagger \inb[0.75cm]{	  2 h_a m_{Q^c}^2
		  				+ 4 h_b m_{\Phi ba}^2
		  				+ 4 m_Q^2 h_a 
		  				}
	\\
	&	& + \; 4 A_{Q a}^\dagger A_{Q a}
		  - \frac{1}{3} M_1 M_1^\dagger g_1^2
		  - 6 M_R M_R^\dagger g_R^2
		  - \frac{32}{3} M_3 M_3^\dagger g_3^2
		  - \eighth g_1^2 \massSum[S]
\end{eqnarray}
%
%
\begin{eqnarray}
\nonumber
16 \pi^2 \deriv{}{t} m_L^2
	& =	&   6 m_L^2 f f^\dagger 
		  + f \inb{	  6 f^\dagger m_L^2
		  		+ 12 \inp{m_L^2}^T f^\dagger
		  		+ 12 f^\dagger m_\Delta^2
		  	    }
	\\ \nonumber
	&	& + \; 2 m_L^2 h_a^\prime h_a^{\prime \, \dagger}
		  + h_a^\prime \inb[0.75cm]{	  2 h_a^{\prime \, \dagger} m_L^2
			  			+ 4 m_{L^c}^2 h_a^{\prime \, \dagger} 
			  			+ 4 h_b^{\prime \, \dagger} m_{\Phi ab}^2
						}
	\\
	&	& + \; 12 A_f A_f^\dagger
		  + 4 A_{L a} A_{L a}^\dagger
		  - 3 M_1 M_1^\dagger g_1^2
		  - 6 M_L M_L^\dagger g_L^2
		  - \frac{3}{8} g_1^2 \massSum[S]
\end{eqnarray}
%
%
%
\begin{eqnarray}
\nonumber
16 \pi^2 \deriv{}{t} m_{L^c}^2
	& =	&   6 m_{L^c}^2 f_c^\dagger f_c 
		  + f_c^\dagger \inb{	  6 f_c m_{L^c}^2 
					+ 12 \inp{m_{L^c}^2}^T f_c
					+ 12 f_c m_{\Delta^c}^2
					}
	\\ \nonumber
	&	& + \; 2 m_{L^c}^2 h_a^{\prime \, \dagger} h_a^\prime
		  + h_a^{\prime \, \dagger} \inb{	  2 h_a^\prime m_{L^c}^2
							+ 4 m_L^2 h_a^\prime
							+ 4 h_b^\prime m_{\Phi ba}^2
						  }
	\\
	&	& + \; 12 A_{f^c}^\dagger A_{f^c} 
		  + 4 A_{L a}^\dagger A_{L a}
		  - 3 M_1 M_1^\dagger g_1^2
		  - 6 M_R M_R^\dagger g_R^2
		  + \frac{3}{8} g_1^2 \massSum[S]
\end{eqnarray}
%
%
\begin{eqnarray}
\nonumber
16 \pi^2 \deriv{}{t} m_{\Delta}^2
	& =	&   \Tr \inb[0.75cm]{	4 f^\dagger f m_\Delta^2
					+ 8 f^\dagger m_L^2 f
				    }
		  + \mu_{\Delta}^{\alpha \, *} \inb{	  2 \mu_\Delta^\alpha m_\Delta^2
		  					+ 2 \mu_\Delta^\alpha m_{\bar{\Delta}}^2
		  					+ 2 \mu_\Delta^\beta \inp{m_S^2}^{\beta \alpha}
		  				     }
	\\
	&	& + \; 4 \Tr \inp{ A_f^\dagger A_f } 
		  + 2 A_\Delta^{\alpha \, *} A_\Delta^\alpha
		  - 12 M_1 M_1^\dagger g_1^2
		  - 16 M_L M_L^\dagger g_L^2
		  + \frac{3}{4} g_1^2 \massSum[S]
\end{eqnarray}
%
%
\begin{eqnarray}
\nonumber
16 \pi^2 \deriv{}{t}  m_{\bar{\Delta}}^2
	& =	&   \mu_\Delta^{\alpha \, *} \inb{	  2 \mu_\Delta^\alpha m_{\bar{\Delta}}^2
							+ 2 \mu_\Delta^\alpha m_{\Delta}^2
							+ 2 \mu_\Delta^\beta \inp{m_S^2}^{\beta \alpha}
						     }
	\\
	&	& 
		  + \; 2 A_\Delta^{\alpha \, *} A_\Delta^\alpha
		  - 12 M_1 M_1^\dagger g_1^2
		  - 16 M_L M_L^\dagger g_L^2
		  - \frac{3}{4} g_1^2 \massSum[S]
\end{eqnarray}
%
%
\begin{eqnarray}
\nonumber
16 \pi^2 \deriv{}{t} m_{\Delta^c}^2
	& =	&   \Tr \inb[0.75cm]{	4 f_c f_c^\dagger m_{\Delta^c}^2
					+ 8 f_c m_{L^c}^2 f_c^\dagger
				    }
		  + \mu_{\Delta^c}^{\alpha \, *} \inb{	  2 \mu_{\Delta^c}^\alpha m_{\Delta^c}^2
		  					+ 2 \mu_{\Delta^c}^\alpha m_{\bar{\Delta}^c}^2
		  					+ 2 \mu_{\Delta^c}^\beta \inp{m_S^2}^{\beta \alpha}
		  				     }
	\\
	&	& + \; 4 \Tr \inp{ A_{f^c} A_{f^c}^\dagger } 
		  + 2 A_{\Delta^c}^{\alpha \, *} A_{\Delta^c}^\alpha
		  - 12 M_1 M_1^\dagger g_1^2
		  - 16 M_R M_R^\dagger g_R^2
		  - \frac{3}{4} g_1^2 \massSum[S]
\end{eqnarray}
%
%
\begin{eqnarray}
\nonumber
16 \pi^2 \deriv{}{t}  m_{\bar{\Delta}^c}^2
	& =	&   \mu_{\Delta^c}^{\alpha \, *} 
			\inb{	  2 \mu_{\Delta^c}^\alpha m_{\bar{\Delta}^c}^2
				+ 2 \mu_{\Delta^c}^\beta \inp{m_S^2}^{\beta \alpha}
				+ 2 \mu_{\Delta^c}^\alpha m_{\Delta^c}^2
			     }
	\\
	&	& 
		  + \; 2 A_{\Delta^c}^{\alpha \, *} A_{\Delta^c}^\alpha
		  - 12 M_1 M_1^\dagger g_1^2
		  - 16 M_R M_R^\dagger g_R^2
		  + \frac{3}{4} g_1^2 \massSum[S]
\end{eqnarray}
%
%
\begin{eqnarray}
\nonumber
16 \pi^2 \deriv{}{t} m_{\Phi ab}^2
	& =	&   m_{\Phi ac}^2 \inb{ \YPhimnYPhimn{c}{b} }
	\\ \nonumber
	&	& 
		  + \inb{ \YPhimnYPhimn{a}{c} } m_{\Phi cb}^2
	\\ \nonumber
	&	& + \Tr \inb{	  6 h_a^\dagger h_b m_{Q^c}^2
		  		+ 6 h_a^\dagger m_Q^2 h_b
				+ 2 h_a^{\prime \, \dagger} h_b^\prime m_{L^c}^2
				+ 2 h_a^{\prime \, \dagger} m_{L}^2 h_b^\prime
				+ 6 A_{Q a}^\dagger A_{Q b} 
				+ 2 A_{L a}^\dagger A_{L b}
			    }
	\\ \nonumber
	&	&
		  + \; 8 \inb{ \mu_\Phi^\alpha m_{\Phi}^2 \mu_\Phi^{\alpha \, \dagger} }_{ba}
		  + 8 \inb{ \mu_\Phi^{\alpha \, \dagger} \mu_\Phi^\beta }_{ab} \inp{m_S^2}^{\beta \alpha}
		  + 8 \inb{ A_\Phi^{\beta \, \dagger} A_\Phi^\beta }_{ab}
	\\
	&	&
		  - \delta_{ab} \inb{ 6 g_L^2 M_L M_L^\dagger + 6 g_R^2 M_R M_R^\dagger }
\end{eqnarray}
%
%
\begin{eqnarray}
\nonumber
16 \pi^2 \deriv{}{t} \inp{m_S^2}^{\alpha \beta}
	& =	&    \inp{m_S^2}^{\alpha \rho} \inb{ \YSmnYSmn{\rho}{\beta} }
	\\ \nonumber
	&	& + \inb{ \YSmnYSmn{\alpha}{\rho} } \inp{m_S^2}^{\rho \beta}
	\\ \nonumber
	&	& + \; 6 \mu_\Delta^{\alpha \, *} \mu_\Delta^\beta m_{\bar{\Delta}}^2
		  + 6 \mu_\Delta^{\alpha \, *} \mu_\Delta^\beta m_{\Delta}^2
		  + 6 \mu_{\Delta^c}^{\alpha \, *} \mu_{\Delta^c}^\beta m_{\bar{\Delta}^c}^2
		  + 6 \mu_{\Delta^c}^{\alpha \, *} \mu_{\Delta^c}^\beta m_{\Delta^c}^2
	\\ \nonumber
	&	& + \; 32 \Tr \inp{ \mu_\Phi^{\alpha \, \dagger} \mu_\Phi^\beta m_{\Phi}^2 }
		  + 2 \conj{\inp{ Y^{\alpha \rho \mu} }} Y^{\beta \rho \nu} \inp{ m_S^2}^{\nu \mu}
	\\
	&	& + \; 6 A_\Delta^{\alpha \, *} A_\Delta^\beta
		  + 6 A_{\Delta^c}^{\alpha \, *} A_{\Delta^c}^\beta
		  + 16 \Tr \inp{ A_\Phi^{\alpha \, \dagger} A_\Phi^\beta }
		  + \conj{\inp{A_S^{\alpha \mu \nu}}} A_S^{\beta \mu \nu}
\end{eqnarray}



\section{Conclusion}

In this paper we have calculated the RGEs to one loop order for two different types of SUSYLR models--one which breaks $SU(2)_R$ via doublets and the other using triplets.  These equations should prove to be useful tools for relating the details of SUSYLR models to observable phenomena, thereby constraining the parameter space and perhaps verifying if SUSYLR models are viable extensions of the standard model.

\section*{Acknowledgments}
We would like to thank Rabindra Mohapatra for his helpful discussions.  We would also like to thank
 Kaladi Babu for contributing materials providing insight into diagram contributions to the RGEs.  This work was supported in part by the National Science Foundation grant PHY-0354401 and the Center for Particle and String Theory.



\end{document}